\documentclass[fleqn,usenatbib]{mnras}
\usepackage[T1]{fontenc}
\usepackage{commath}
\usepackage{amsmath}	
\usepackage{amssymb}	
\usepackage{ae,aecompl}
\usepackage{graphicx}	
\usepackage[usenames,dvipsnames,svgnames,table]{xcolor}
\usepackage[table]{xcolor}
\usepackage{verbatim}
\usepackage[font = small, labelfont=bf, labelsep=period]{caption}
\usepackage{subcaption}
\usepackage{breqn}
\usepackage{float}
\usepackage{dblfloatfix}

\usepackage[utf8]{inputenc}
\usepackage{multicol}
\usepackage{bibspacing}
\usepackage{lipsum}
\usepackage{hyperref}
\usepackage{orcidlink}
\usepackage{diagbox}
\usepackage[inline]{enumitem}
\usepackage{enumitem}
\usepackage[normalem]{ulem}
\setlist{
	listparindent=\parindent,
	parsep=0pt,
}

\hypersetup{ %
			colorlinks=true,%
			pdfauthor={Magdalena Siwek},%
			pdftitle={calypso},%
			citecolor=blue%
			}

\newcommand{\msol}{M_\odot}

\newcommand{\Mb}{$M_{\rm b}$}
\newcommand{\Mbd}{$\dot{M}_{\rm b}$}
\newcommand{\qb}{$q_{\rm b}$}
\newcommand{\eb}{$e_{\rm b}$}

\newcommand{\ab}{$a_{\rm b}$}

\newcommand{\calypso}{\texttt{calypso}}

\title[calypso]{\texttt{calypso}: 
a Parameter-Conditioned Stochastic Surrogate Model 
for Circumbinary Accretion Time-Series}
\author[Magdalena Siwek et al.]{
Magdalena Siwek\,\orcidlink{0000-0002-1530-9778},$^{1,2}$\thanks{E-mail: mss931@nyu.edu}, 
Matt Ho\,\orcidlink{0000-0003-3207-8868}$^{1}$, 
Earl Bellinger\,\orcidlink{0000-0003-4456-4863}$^{3,4}$
\\
$^{1}$New York University, Center for Cosmology and Particle Physics, 726 Broadway, New York, NY 10003, USA\\
$^{2}$Columbia University, Pupin Hall, 538 W 120th St, New York, NY 10027, USA\\
$^{3}$Department of Astronomy, Yale University, New Haven, CT 06511, USA\\
$^{4}$Institute for Foundations of Data Science, Yale University, New Haven, CT 06511, USA\\
}

\date{Accepted XXX. Received YYY; in original form ZZZ}

\pubyear{2026}

\begin{document}
\label{firstpage}
\pagerange{\pageref{firstpage}--\pageref{lastpage}}
\maketitle

\begin{abstract}
    We present \calypso, a parameter-conditioned stochastic surrogate model 
	for circumbinary accretion flows. We represent the total and individual 
	accretion time series in a PCA basis and 
	model the resulting coefficients as draws 
	from a multivariate Gaussian distribution over the latent PCA coefficients.
	We specifically include the aleatoric uncertainty
	of the time series in the model, enabling the emulator to capture the 
	inherent stochasticity of the accretion process and the 
	long-term modulation due to disk precession. 
	We further explore the epistemic uncertainty in the model due to 
	limited training data and interpolation in the (\eb,\qb) parameter space
	and find that the data does not support inclusion of this added variance term.
	We present the properties of existing simulation suites of circumbinary accretion, 
	and run new simulations to fill in gaps in the parameter space, as well as 
	a set of 13 test simulations for validation of the emulator.
	We publish \calypso \ as a pip-installable Python package with
	an open-source codebase and comprehensive documentation,
	and demonstrate use-cases for current and upcoming transient surveys.
	We additionally derive a closed-form Gaussian likelihood that
	enables direct inference of (\eb, \qb) from observed
	accretion-rate time series.
\end{abstract}


\section{Introduction}
A growing body of evidence from multi-messenger observations suggests that 
(super-)massive black hole binaries (MBHBs) 
form binary systems in post-merger galaxies. 
Although none are observationally confirmed, 
recent results from Pulsar Timing Array (PTA) collaborations 
indicate tantalizing evidence
for a gravitational wave background (GWB) signal 
\citep{NANOGravDetection2023,EPTADetection2023,PPTADetection2023,CPTADetection2023}, 
likely originating from the continuous inspiral of 
MBHBs across cosmic time \citep{NANOGravMBHBs2023}.
Additionally, a growing number of candidate MBHBs 
are identified in transient surveys 
(e.g., \citealt{Graham2015a, Graham2015b, Charisi2016, Liu2019}), 
though identifying false positives remains a challenge due to intrinsic quasi-periodic variability in 
active galactic nuclei (AGN). 
Most promising is the
Vera C. Rubin Observatory Legacy Survey of Space and Time (LSST; \citealt{LSST2019}),
which will provide a wealth of data on the optical and near-infrared sky,
and result in the most comprehensive survey of the MBHB population.

The identification of MBHBs in transient surveys, such as LSST, relies
on the detection of periodic variability in the 
light curves of AGN. 
The variability is thought to arise from the modulation 
of the accretion rates onto the binary components by the 
circumbinary disk (CBD) surrounding the binary.
Hydrodynamic simulations of circumbinary accretion 
have shown that the accretion rates onto the 
binary components exhibit variability on 
timescales comparable to the binary orbital period 
(e.g., \citealt{MacFadyen2008,Farris2014,Shi2012,Moody2019,Cocchiararo2024}), 
as well as long-term modulation on timescales much 
longer than the binary orbital period due to the precession of the CBD 
\citep{Munoz2016,Miranda2016,Munoz2020,Ragusa2020,Siwek2023a}.

So far, however, searches for periodic variability in AGN
lightcurves have mostly assumed that this variability 
is periodic and quasi-sinusoidal (e.g., \citealt{Graham2015b, Xin2026, ElBadry2026}).
This is a significant simplification, 
as hydrodynamic simulations have shown that the accretion variability is
highly non-sinusoidal, and exhibits a range of morphologies across the binary parameter space.
Specifically, eccentric binaries tend to exhibit more bursty, 
non-sinusoidal variability
(e.g., \citealt{Zrake2021, DOrazio2024}).
It is possible that this simplification has led to the non-detection 
of MBHB candidates in transient surveys (e.g., \citealt{Lin2026, Park2026}),
or the misclassification of true MBHB candidates as 
false positives due to their non-sinusoidal variability signatures.
Future searches for MBHB candidates in transient 
surveys will therefore require more realistic models of the accretion variability,
which can capture the non-sinusoidal nature of the 
variability and its dependence on the binary parameters.

Here we present a systematic study of the accretion variability across
the eccentricity (\eb) and mass ratio (\qb) parameter space, and introduce \calypso, 
a parameter-conditioned stochastic surrogate model for circumbinary accretion time series.
\calypso \ stands for `Circumbinary Accretion Lightcurves Yielded via Probabilistic Spectral Operators',
and is designed to predict the total, primary and secondary accretion rates
of an accreting binary system with arbitrary eccentricity \eb\ and mass ratio \qb.
We train \calypso \ on a suite of hydrodynamic simulations of circumbinary accretion,
and show that it can capture the variability of the accretion rates across the parameter space,
showing that accretion is largely dominated by non-sinusoidal variability, 
particularly for eccentric binaries.
We find that many binaries in the (\eb,\qb) parameter space 
exhibit double-peaked, bursty, and saw-toothed variability, 
in contrast with sinusoidal variability.
\calypso \ is fully stochastic, returning samples from a 
multivariate Gaussian distribution over the latent model coefficients,
thus providing a principled way to capture the uncertainty in the accretion variability.
The stochastic nature of \calypso \ will also be useful for inference of
binary parameters from observed light curves, and therefore 
for population studies of MBHBs in the era of LSST and PTA detections.

This paper is structured as follows.
In \S\ref{sec:methods},
we describe our methods, beginning with the 
hydrodynamic simulations of circumbinary accretion 
and the properties of the resulting accretion time series.
We provide an in-depth description of the data pre-processing and training of \calypso \ via 
Principal Component Analysis (PCA) on the accretion time series,
and the variance model we employ to capture the 
aleatoric and epistemic uncertainty of the emulator.
We also describe the interpolation scheme we use to 
enable predictions for arbitrary (\eb,\qb) pairs 
within the parameter space covered by our training data.
In \S\ref{sec:validation}, we present the validation of
\calypso \ on a set of 13 test simulations,
and use the validation results to determine the optimal 
number of PCA components to retain in the final emulator.
We also assess the performance of the emulator when 
including the epistemic variance term, showing that the data does
not support the inclusion of this added variance term.
In \S\ref{sec:evaluation}, we specifically
evaluate the performance of \calypso \ in 
the (\eb,\qb) parameter space, showing the emulator performance
directly by comparing time series generated by \calypso \ to the
corresponding hydrodynamic simulations,
and compare the variance estimates from the emulator to 
the true variance in the test simulations.
We show our method to convert \calypso \ accretion rates into
luminosities and apparent magnitudes in LSST bands, 
and present example light curves for a subset of test simulations 
in \S\ref{sec:lsst_application}.
In \S\ref{sec:discussion}, we summarize and discuss
our results, the limitations of our current model, 
and outline future directions for improving \calypso.
In \S\ref{sec:installation} we provide instructions for installation and use of \calypso.
In Appendix~\ref{sec:appendix_likelihood} we derive a closed-form Gaussian likelihood for (\eb, \qb) inference from observed accretion-rate time series.

\section{Methods}

\label{sec:methods}

\subsection{Hydrodynamic Simulations of Circumbinary Accretion}
\label{sec:hydro}
This work is based on a suite of hydrodynamic simulations of circumbinary accretion disks 
described in \cite{Siwek2023a, Siwek2023b}, 
as well as new simulations to fill in the parameter space, run with the same methods.
We refer the reader to these papers for an in-depth
description of the simulation setup, but briefly summarize the key points here.
The simulations were performed using the \texttt{Arepo} code \citep{Springel2010, Pakmor2016} 
in its Navier-Stokes version \citep{Munoz2013}. 
The simulations include a central binary represented by two sink particles,
which move on fixed Keplerian orbits with eccentricity \eb\ and mass ratio \qb.
The sink particles accrete gas from the surrounding disk, 
and the accretion rates onto each sink are recorded on-the-fly.
The system is scale-free, with the 
binary mass and semimajor axis set to unity, 
and the binary orbital period $P_{\rm b} = 2 \pi$.
The accretion disk is modeled as a viscous, locally isothermal fluid with an 
$\alpha$-disk viscosity prescription, similar
to previous work in the literature 
(e.g., \citealt{Farris2014}, \citealt{Munoz2019}, \citealt{Tiede2020}).
The hydrodynamic parameters of the disk include $\alpha = 0.1$ and a disk aspect
ratio of $H/R = 0.1$.
While keeping these parameters constant, a large parameter study over binary mass ratio 
\qb\ and eccentricity \eb\ was performed and run for $10,000\, P_{\rm b}$ 
to ensure that the accretion rates have reached a quasi-steady state. 
The accretion time series is then extracted for the last 5000 binary orbits, 
yielding a total of 5000 data points per time series at a cadence of 100 points per binary orbit,
forming the basis of our training dataset for \calypso.

This work contains previously published data from \cite{Siwek2023a, Siwek2023b} covering the
original parameter space, with some additional simulations to fill in gaps, 
most notably for \eb $= 0.7$. In total, that brings the 
number of simulations used for training to 100, 
with eccentricities $e_{\rm b} \in [0, 0.05, 0.1, 0.2, 0.3, 0.4, 0.5, 0.6, 0.7, 0.8]$ 
and mass ratios $q_{\rm b} \in [0.1, 0.2, 0.3, 0.4, 0.5, 0.6, 0.7, 0.8, 0.9, 1.0]$.
Additionally, 13 test cases were produced for validation and testing of the emulator. 
The distribution of the test cases in the parameter space is 
shown in Figure \ref{fig:pspace_test}.

\subsubsection{Properties of the accretion time series}
\label{sec:timeseries}
Here we discuss the properties of the accretion time series obtained from the simulations.
The accretion rates for each sink are obtained by 
summing the mass flux onto each sink particle in the simulation
on-the-fly, and written into an output file at a cadence of 100 points per binary orbit.
In this work we analyze the accretion time series over 5000 binary orbits, yielding enough 
data to assess the long-term variability and stability of the accretion rates.

Simulations have long shown that accretion rates exhibit variability 
on timescales comparable to the binary orbital period (e.g., \citealt{MacFadyen2008,Farris2014}).
Here, we analyze this short-term variability for our entire simulation suite
over eccentricity \eb \ and mass ratio \qb.
We characterize the accretion variability frequency across our 
simulation suite using Gaussian Process regression 
on the power spectrum of each window.
Specifically, we slice the 5000 binary orbit time series per binary into 500 windows, 
yielding 10 binary orbits per window. 
We then interpolate each window to ensure even 
time-spacing, and perform a fast Fourier transform to obtain the power spectrum of 
the accretion variability in each window. 
We note that the interpolation does not affect the variability signature,
as the original time-series data is largely evenly spaced, with only occasional
binning irregularities due to the tree-based nature of the \texttt{Arepo} code. 
Our interpolation scheme corrects for these irregularities, 
ensuring that the time-series data is uniformly spaced across all windows,
but does not meaningfully change the number of bins per binary orbit or the underlying variability signature.

We model the $\log_{10}$ of the power spectrum with a Gaussian Process (GP).
We choose a Mat\'{e}rn $\nu = 0.5$ covariance kernel for its ability to 
capture sharp peaks in power spectra, and a prior 
mean function centered at a frequency $\Omega/\Omega_{\rm b} = 1$ 
motivated by the fundamental frequency of the binary. 

\begin{figure*}
	\centering
	\includegraphics[width=\textwidth]{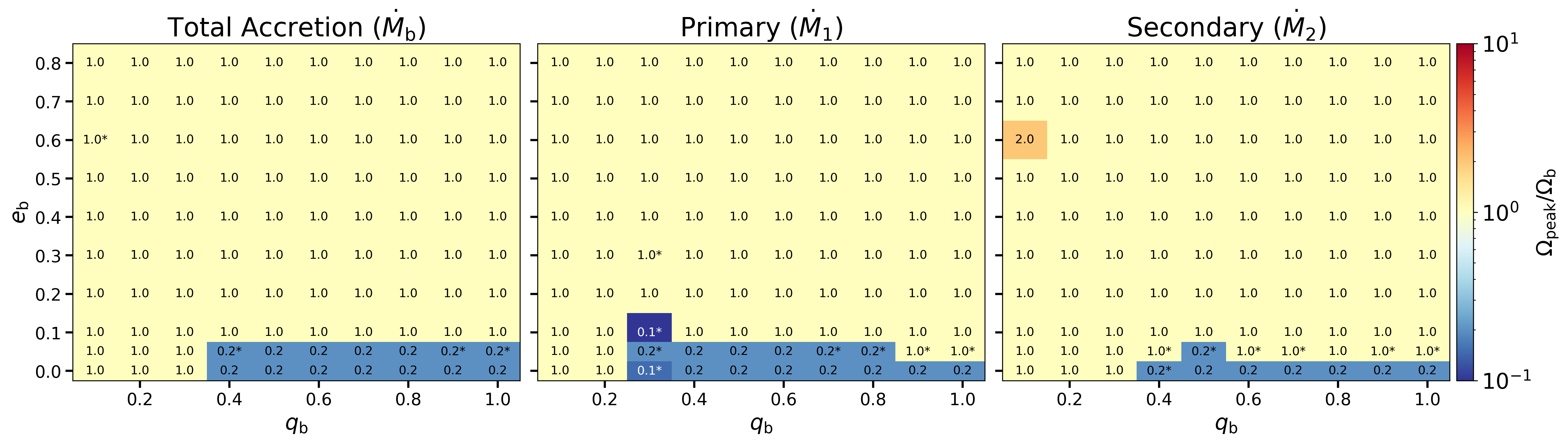}
	\caption{Peak frequency of the accretion-variability power spectrum
		across the training grid, shown for the binary-total
		$\dot M_{\rm b}$ (left), the primary $\dot M_1$ (centre), and
		the secondary $\dot M_2$ (right). Each cell corresponds to one
		$(e_{\rm b}, q_{\rm b})$ training pair; the colour and the
		in-cell number give the dominant mixture-posterior peak in
		units of the binary orbital frequency $\Omega_{\rm b}$
		(log colour scale centred on
		$\Omega_{\rm peak}/\Omega_{\rm b}=1$). Cells whose mixture
		posterior across the 500 windows is multimodal are marked with
		an asterisk and shown in detail in
		Figure~\ref{fig:peak_multimodal}. Most binaries peak at the
		fundamental $\Omega_{\rm b}$. Some components show slightly
		higher peak frequencies, while higher-$q_{\rm b}$,
		near-circular binaries show a peak near
		$\Omega/\Omega_{\rm b} \simeq 0.2$ corresponding to the
		well-known $\sim 5\,P_{\rm b}$ ``lump'' oscillation
		\citep[e.g.,][]{Farris2014}.}
	\label{fig:peak_frequency}
\end{figure*}

For each window, we collect the peak frequency of the power spectrum 
from the GP posterior, along with the full-width at half max (FWHM) 
of the peak as a measure of uncertainty.
The FWHM is then converted into a standard deviation by dividing by $2 \sqrt{2 \ln 2}$,
assuming a Gaussian shape for the peak.
Each window therefore contributes a Gaussian distribution over the peak frequency, 
\begin{equation}
	p(\Omega) = \frac{1}{\sqrt{2 \pi \sigma^2}} \exp\left(-\frac{(\Omega - \mu)^2}{2 \sigma^2}\right).
\end{equation}
The mean PDF over all windows is then given by the mixture of Gaussians,
\begin{equation}
	p_{\rm mixture}(\Omega) = \frac{1}{N} \sum_{i=1}^{N} p_i(\Omega),
	\label{eq:mixture}
\end{equation}
where $N=500$ is the number of windows.
The highest peak of the mixture probability density function (PDF) 
is identified as the dominant mode of the distribution.
Each cell in Figure \ref{fig:peak_frequency} shows the
dominant mode we identify. 
In most cases, the distribution of peak frequencies is unimodal and well-constrained,
and the peak frequency is close to the fundamental frequency of the binary, $\Omega_{\rm b}$.
In cases where there is clear multi-modality in the distribution of peak frequencies 
across the 500 windows,
we indicate this by adding an asterisk (*) to the cell.

\begin{figure*}
	\centering
	\includegraphics[width=\textwidth]{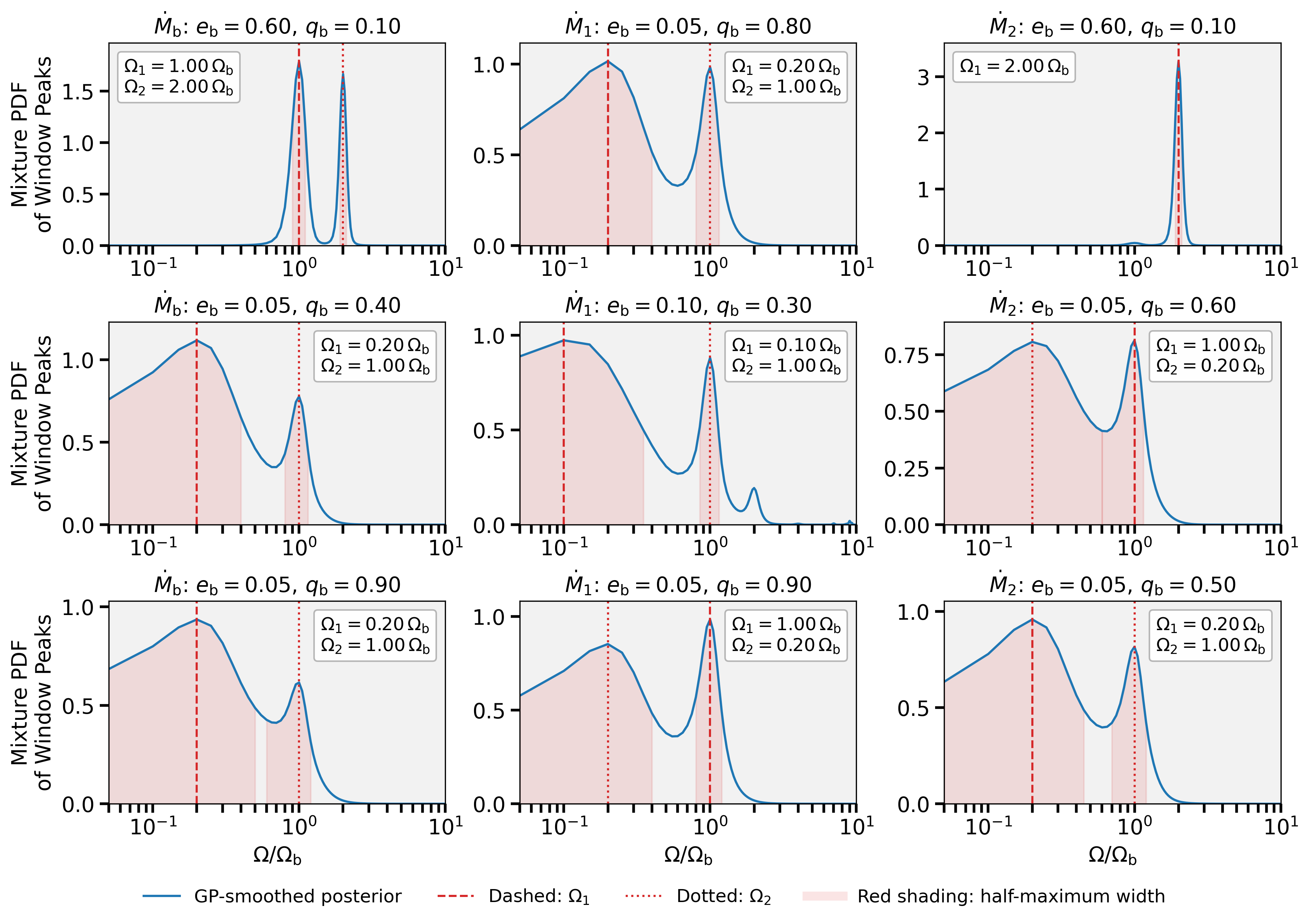}
	\caption{Mixture-posterior peak-frequency distributions
		(equation~\ref{eq:mixture}, with $N=500$ windows per binary)
		for representative multimodal training cases.
		Columns correspond to the binary-total $\dot M_{\rm b}$ (left),
		primary $\dot M_1$ (centre), and secondary $\dot M_2$ (right).
		For $\dot M_{\rm b}$ and $\dot M_1$, the three rows are the
		multimodal cases with the largest secondary-to-primary
		posterior peak-height ratio. For $\dot M_2$, the first row
		is the outlier case $(e_{\rm b}, q_{\rm b}) = (0.6, 0.1)$,
		whose dominant peak lies far from $\Omega_{\rm b}$, and the
		next two rows are the top two by secondary-to-primary ratio.
		Within each panel, the solid curve is the GP-smoothed
		mixture PDF\@; the dashed and dotted vertical lines mark the
		locations of the dominant ($\Omega_1$) and secondary
		($\Omega_2$) peaks; the red shading indicates the
		full-width at half-maximum of the dominant peak. The
		$(e_{\rm b}, q_{\rm b})$ pair labels each panel. Both axes
		are logarithmic.}
	\label{fig:peak_multimodal}
\end{figure*}

In Figure \ref{fig:peak_multimodal}, we show the mixture PDFs of some binaries 
with either clear multimodality or an outlier peak (e.g., $\dot{M}_2$ with \eb=0.6, \qb=0.1)
for all three components \Mbd, $\dot{M}_1$, and $\dot{M}_2$.
Typically, the primary peak is at the fundamental frequency of the binary, $\Omega_{\rm b}$,
with an accompanying secondary peak at $\Omega=0.2$. This is related
to the orbital frequency of an overdensity (``lump'') at the inner edge of the circumbinary disk, 
which is a well-known feature of near-equal mass, 
near-circular binaries (e.g., \citealt{Farris2014}), 
and has also been observed in MHD simulations (e.g., \citealt{Shi2012}).

More recently, hydrodynamic simulations have also indicated 
that there is a long-term 
modulation of the accretion rates on timescales much longer than the 
binary orbital period (e.g., \citealt{DOrazio2013, Munoz2019, Siwek2023a}). 
The long-term modulations are 
typically on timescales of several hundreds of binary orbits, 
and associated with the precession timescale of the CBD \citep{Siwek2023a}. 
However, the existence, strength and period of this modulation all
depend on the binary parameters \eb \ and \qb, with some CBDs
`locking' at a fixed pericenter angle rather than precessing.
These systems show little to no long-term modulation (\citealt{Siwek2023a}).

Precession and long-term modulation complicate the 
modeling of short-term variability in the accretion 
rates by introducing uncertainty in the short-timescale 
variability signature. This effect is particularly pronounced 
for binaries with high mass ratios and eccentricities, 
which exhibit both higher-amplitude variability 
and more pronounced CBD precession. 
We interpret the precession-driven modulation as a significant 
additional source of stochasticity in the accretion signal, 
which can in some regimes dominate over the intrinsic variability of the flow. 
The intrinsic stochasticity of the accretion flow—arising from 
intermittent stream penetration into the cavity due to perturbations 
in the viscous-gravitational torque balance at the cavity's 
inner edge—provides a baseline level of variability upon 
which this precession-driven modulation is superimposed.
The quantification of these two sources of uncertainty in the characteristic accretion 
variability for any binary in our parameter space is a central feature of \calypso,
the fundamental methods of which we describe in the following sections.

Building on the simulation suite described above, \calypso\ emulates
the total, primary, and secondary accretion rates of a binary at arbitrary
(\eb,\qb) within the trained box. The remainder of this section describes
the pre-processing applied to the raw accretion time series
(\S\ref{sec:data_preprocessing}), the PCA-based training of the emulator
(\S\ref{sec:training}), the (\eb,\qb) interpolation scheme
(\S\ref{sec:interpolation}), and the variance model used to capture
the aleatoric and epistemic uncertainty
(\S\ref{sec:uncertainty}).

\subsection{Data Pre-processing}
\label{sec:data_preprocessing}
Specifically, we create a training dataset that consists of 500 time-series examples (`windows') 
showing 10 $P_{\rm b}$ for each (\eb,\qb) pair in our training dataset. 
Each time series is interpolated prior to training to ensure that timesteps are uniform across all windows,
with negligible effect on the underlying variability.

Due to the long-term precession of the circumbinary disk, 
the accretion variability
exhibits a long-timescale drift. 
Since we are training on relatively short windows of 10 $P_{\rm b}$, 
the long-timescale drift due to disk precession is not captured in each individual window. 
However, stacking the entire set of raw 10 $P_{\rm b}$ windows for each (\eb,\qb) pair would
average out some of the short-term signals due to the long-timescale drift. 
This is particularly noticeable
for binaries with high mass ratios and near-circular orbits, 
which exhibit lower-amplitude variability
and peak variability frequencies corresponding to 
$\sim 5\, P_{\rm b}$ (see Figure \ref{fig:peak_frequency}).

We therefore pre-process the time-series data so that each window is detrended, 
removing the long-timescale drift due to disk precession in each window. 
Specifically, we find the peak frequency of each binary (shown in 
Figure \ref{fig:peak_frequency}),
and find the associated phase $\phi_{\rm peak}$ of the Fourier component at that frequency,
\begin{equation}
	\phi_{\rm peak} = \arctan\left(\frac{\text{Im}(\tilde{X}(\Omega_{\rm peak}))}{\text{Re}(\tilde{X}(\Omega_{\rm peak}))}\right).
\end{equation}
We then shift the time-series data in each window by the corresponding phase,
so that the variability peaks are aligned across all windows for each (\eb,\qb) pair.
Specifically, the time series becomes,
\begin{equation}
	\dot{M}(t) \rightarrow \dot{M}(t + \phi_{\rm peak}/\Omega_{\rm peak}),
\end{equation}
where $\phi$ is the phase of the Fourier component at the peak frequency $\Omega_{\rm peak}$.
This allows us to capture the short-term variability in 
each window without averaging out the signal due to 
the long-term drift. We point out that transient surveys are phase-agnostic 
to the long-term variability of the CBD, 
since they will only observe a small fraction of the precession period.

The interpolated, de-trended time series of \Mbd, $\dot{M}_1$ and $\dot{M}_2$ are
concatenated to form a single feature vector for each window.
This dataset forms a matrix where each row corresponds to a single window and each column corresponds
to a time point in the concatenated \Mbd, $\dot{M}_1$ and $\dot{M}_2$ time series.
Specifically, we end up with a matrix of shape $(N \times m, 3 \times T)$, 
where $N = 500$ is the number of windows per (\eb,\qb) pair and $m=100$ is 
the number of (\eb,\qb) pairs in the training set. 
$T = n_{\rm bins} \times n_{\rm orb}$ is the number of timesteps in each window. 

\subsection{Training}
\label{sec:training}
Following data pre-processing described in \S\ref{sec:data_preprocessing},
we obtain a global PCA basis that captures 
the variability across the accretion time series over our entire (\eb,\qb) parameter space. 
The matrix $X \in \mathbb{R}^{M \times 3T}$ is 
the training data matrix, with the number of rows corresponding to the number of samples ($ M = N \times m$) 
and the number of columns corresponding to the number of timesteps in the concatenated 
\Mbd, $\dot{M}_1$, and $\dot{M}_2$ time series (number of `features').
We first mean-center the data by subtracting the mean of each column 
from the corresponding entries in that column,
ensuring that the PCA optimally captures the variance in the data.

We perform a singular value decomposition (SVD) of the mean-centered data matrix $X$,
\begin{equation}
	X = U \Sigma V^T,
\end{equation}
where the columns of $V$ are the PCA basis vectors, the columns of $U \Sigma$ are the PCA coefficients for each window,
and $\Sigma$ is a diagonal matrix containing the singular values, 
which are related to the explained variance of the corresponding PCA components.
Specifically, $\Sigma$ is diagonal and ordered, such that
\begin{equation}
	\Sigma_{11} \geq \Sigma_{22} \geq ... \geq \Sigma_{kk} \geq 0,
\end{equation}
and related to the eigenvalues of the covariance matrix of $X$ by 

\begin{equation}
	\lambda_k = \frac{\Sigma_{kk}^2}{(N m - 1)}.
\end{equation}

The explained variance ratio of the $k$-th PCA component is then given by
\begin{equation}
	\tilde{\lambda}_k = \frac{\lambda_k}{\sum_{j=1}^{3 \times T} \lambda_j},
\end{equation}
which represents the proportion of the total variance in the data that is 
captured by the $k$-th PCA component.

\begin{figure}
	\centering
	\includegraphics[width=\columnwidth]{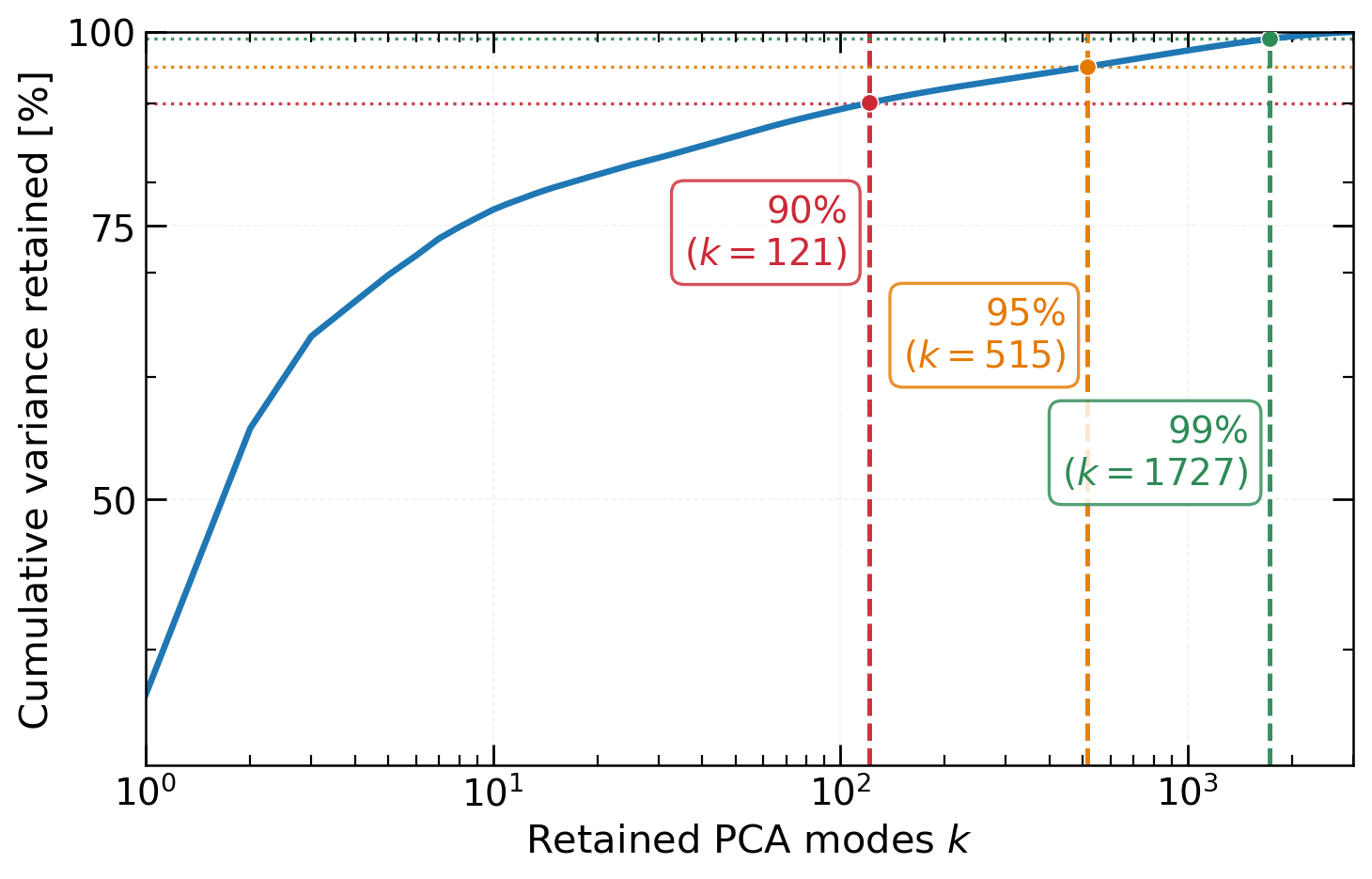}
	\caption{Cumulative explained variance of the joint PCA basis
			(over $\dot M_{\rm b}$, $\dot M_1$, $\dot M_2$) as a function of the
			number of retained principal components $k$, on log-log axes.
			The dotted horizontal lines mark the $90\%$ (red), $95\%$ (orange), and $99\%$ (green)
			variance-retained levels; the dashed vertical lines and annotated
			markers indicate the $k$ at which each threshold is reached.
			This convergence behaviour motivates the search range over $k$
			performed in Figure~\ref{fig:k_sweep}, where the final value
			$k_{\rm final}=142$ is selected from the test-set zMSE.}
	\label{fig:variance_convergence}
\end{figure}

In Figure \ref{fig:variance_convergence}, we show the convergence of the 
explained variance as a function of the number of PCA components.
Specifically, we plot the cumulative explained variance ratio,
\begin{equation}
	\tilde{\lambda}_{\rm cumulative}(k) = \sum_{j=1}^{k} \tilde{\lambda}_j,
\end{equation}
which represents the total proportion of variance captured by the first $k$ PCA components.
We note that the cumulative variance retained scales approximately as $
\propto \log(k)^\gamma$ with $\gamma$ between 1 and 2,
indicating that much of the variance in the data can be captured by a relatively small number of PCA components.
We find that the explained variance converges to $\gtrsim 95\%$ with $\gtrsim 500$ PCA components,
which is a significant dimensionality reduction from the original feature space of $3 \times T (= 3000)$ dimensions.
In \S\ref{sec:validation}, we explain how our test dataset of 13 (\eb,\qb) pairs, which was not included in training,
is used to determine the precise number of PCA components to retain in the final emulator, 
and to validate its performance in the (\eb,\qb) parameter space.

After performing PCA, we obtain a set of PCA basis vectors (the columns of $V$) 
and the corresponding PCA coefficients (the columns of $U \Sigma$) 
for each window in our training dataset.
By including multiple time windows for each (\eb,\qb) pair in the training dataset, 
we capture the stochasticity in the accretion variability.  
Specifically, for each (\eb,\qb) pair, we have $N$ windows, each with $k$ PCA coefficients corresponding to the first $k$ PCA components.
This gives us a matrix of PCA coefficients for each (\eb,\qb) pair,

\begin{equation}
C_{(e_{\rm b}, q_{\rm b})} =
\begin{bmatrix}
c_{11} & c_{12} & \cdots & c_{1k} \\
c_{21} & c_{22} & \cdots & c_{2k} \\
\vdots & \vdots & \ddots & \vdots \\
c_{M1} & c_{M2} & \cdots & c_{Mk}
\end{bmatrix}
\in \mathbb{R}^{N \times k}.
\end{equation}

For each (\eb,\qb) pair, we model the distribution of these coefficient vectors across the 
N windows by estimating a mean vector and covariance matrix,
\begin{equation}
\boldsymbol{\mu} = \frac{1}{N} \sum_{i=1}^{N} \mathbf{c}_i,
\label{eq:mean_vector}
\end{equation}

\begin{equation}
\Sigma = \frac{1}{N - 1} \sum_{i=1}^{N} (\mathbf{c}_i - \boldsymbol{\mu})(\mathbf{c}_i - \boldsymbol{\mu})^\top.
\label{eq:covariance_aleatoric}
\end{equation}

We note that the covariance matrix in our model is dense, but weak except for the first $k=20$ terms,
which show strong covariance between the PCA coefficients. 
Further versions of \calypso \ may want to improve the
speed of the interpolation by using a low-rank approximation of the covariance matrix.
Using these estimated means and covariances, 
we model the PCA coefficient vector for each 
(\eb,\qb) pair as a draw from a k-dimensional multivariate Gaussian distribution,
\begin{equation}
p(\mathbf{c} \mid e_{\rm b}, q_{\rm b}) =
\frac{1}{(2\pi)^{k/2} |\Sigma|^{1/2}}
\exp\left[
-\frac{1}{2}
(\mathbf{c} - \boldsymbol{\mu})^\top
\Sigma^{-1}
(\mathbf{c} - \boldsymbol{\mu})
\right],
\end{equation}
where $\boldsymbol{\mu}$ and $\Sigma$ are the mean vector and 
covariance matrix of the PCA coefficients estimated 
from the M training windows at fixed (\eb,\qb).

\subsection{Interpolation in the (\eb,\qb) parameter space}
\label{sec:interpolation}
To enable predictions for arbitrary (\eb,\qb) pairs within the parameter space covered by our
training data, we perform interpolation of the mean vectors and 
covariance matrices of the PCA coefficients across the (\eb,\qb) parameter space.
However, the interpolation of covariance matrices is 
non-trivial due to the requirement that they must be positive semi-definite.
To address this, we first perform the Cholesky decomposition 
of the covariance matrices. 
Specifically, for each (\eb,\qb) pair, we perform a Cholesky decomposition of the covariance matrix,
\begin{equation}
\Sigma = L L^T,
\end{equation}
where $L$ is a lower triangular matrix.
We then interpolate the mean vectors $\boldsymbol{\mu}$
and the Cholesky factors instead of the covariance matrices directly,
before reconstructing the covariance matrices 
from the interpolated Cholesky factors.
This yields a new multivariate Gaussian distribution 
at the target (\eb,\qb) parameter point. 
We choose to use a simple linear interpolation scheme for both the 
mean vectors and the Cholesky factors
to preserve numerical stability.

\subsection{Aleatoric and Epistemic Uncertainty}
\label{sec:uncertainty}
In our model, the aleatoric uncertainty is captured by the covariance matrix 
$\Sigma$ of the multivariate Gaussian distribution over the PCA coefficients.
This covariance matrix captures the intrinsic variability in the accretion time series for a given (\eb,\qb) pair,
which arises from the stochastic nature of the accretion process and the 
long-term modulation due to disk precession.
The epistemic uncertainty, on the other hand, arises from the 
limited number of training samples and the interpolation process in the (\eb,\qb) parameter space.
We capture the epistemic uncertainty as follows.
For each (\eb,\qb) pair in the training dataset, we compute a mean vector 
of the PCA coefficients across the N windows.
When a new (\eb,\qb) pair is requested through the emulator, 
we perform interpolation of the mean vectors and covariance 
matrices across the training dataset to obtain the mean vector 
and covariance matrix for the new (\eb,\qb) pair.

The epistemic variance measures the spatial variation of the first moment
 (i.e., the mean PCA coefficient vector)
as a function of (\eb,\qb). To do so, we first define a distance-weighted 
mean PCA coefficient vector, with weights between the sampling point and the i-th 
training point defined as follows,
\begin{equation}
	w_i = \frac{1}{(d_i + \epsilon)^\alpha},
\end{equation}
where $d_i$ is the distance in the (\eb,\qb) parameter space between the new point 
and the $i$-th training point,
$\epsilon$ is a small constant to prevent division by zero, and $\alpha$ 
is a hyperparameter that controls the rate of decay of the weights with distance. 
We fiducially set $\epsilon = 10^{-8}$ and $\alpha = 2$, 
which gives more weight to training points that are closer in the parameter space.
The weights are then normalized to sum to unity,
\begin{equation}
	\tilde{w}_i = \frac{w_i}{\sum_{j=1}^{m} w_j}.
\end{equation}
The distance-weighted mean PCA coefficient vector is then given by
\begin{equation}
\boldsymbol{\mu}_{\rm weighted} = \sum_{i=1}^{m} \tilde{w}_i \boldsymbol{\mu}_i,
\end{equation}
where $\boldsymbol{\mu}_i$ is the mean PCA coefficient vector for the $i$-th training point.
The epistemic variance is then defined as the weighted variance of the mean PCA coefficient vectors across the training dataset,
\begin{equation}
	\Sigma_{\rm epistemic} = \sum_{i=1}^m \tilde{w}_i (\boldsymbol{\mu}_i - \boldsymbol{\mu}_{\rm weighted})(\boldsymbol{\mu}_i - \boldsymbol{\mu}_{\rm weighted})^\top,
	\label{eq:covariance_epistemic}
\end{equation}
which captures the uncertainty in the mean PCA coefficient vector due to interpolation in the parameter space
and quantifies the spatial variation of the mean PCA coefficient vector as a function of (\eb,\qb).
We note that this formulation is a first-order approximation that 
captures uncertainty in the interpolated
mean coefficient vector due to spatial variation in the mean coefficient 
vector across the training dataset; uncertainty in the interpolated 
covariance matrix is not separately modeled.
Since aleatoric and epistemic variances are independent,
the final variance used in our emulator is their sum,
\begin{equation}
	\Sigma_{\rm total} = \Sigma_{\rm aleatoric} + \Sigma_{\rm epistemic},
\end{equation}
where $\Sigma_{\rm aleatoric}$ is the covariance matrix of the 
multivariate Gaussian distribution over the PCA coefficients 
for a given (\eb,\qb) pair (see equation \ref{eq:covariance_aleatoric}), and $\Sigma_{\rm epistemic}$ 
is the epistemic variance defined above (equation \ref{eq:covariance_epistemic}).
In \S\ref{sec:validation}, we show the effect of including
the epistemic uncertainty on the emulator fidelity on the test set.
\section{Model Selection and Validation}
\label{sec:validation}
\begin{figure}
	\centering
	\includegraphics[width=\columnwidth]{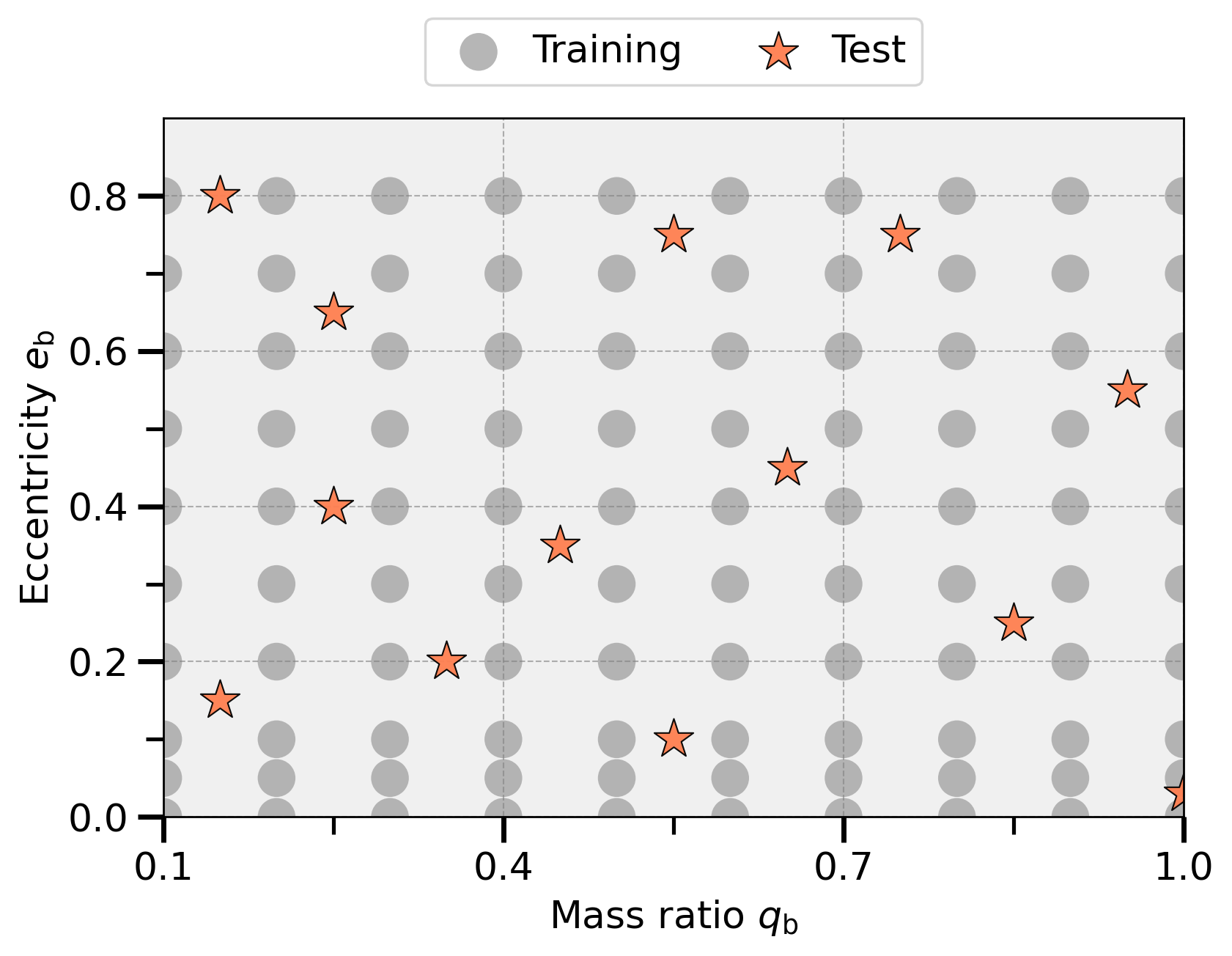}
	\caption{Distribution of the training and test sets in the (\eb,\qb) parameter space.
	The training set consists of 100 (\eb,\qb) pairs shown in grey full circles, 
	while the test set consists of 13 (\eb,\qb) pairs shown in orange stars. 
The test set is designed to evenly cover the parameter space of the training data,
enabling us to validate the performance of \calypso \ across the entire parameter space.}
	\label{fig:pspace_test}
\end{figure}

Following the training and interpolation steps described in
\S\ref{sec:training}~and~\S\ref{sec:interpolation}, we validate the
performance of \calypso \ using a test dataset of 13 (\eb,\qb) 
pairs that were not included in the training process. 
In Figure \ref{fig:pspace_test}, we show the distribution of 
the test set in the (\eb,\qb) parameter space. 
Due to the computational cost of the hydrodynamic simulations, 
we are limited in the number of test cases we can produce, and 
therefore we strategically select test cases that evenly cover 
the parameter space of the training data.
Our test cases were picked by hand to ensure that they are evenly 
distributed across the parameter space, rather than using 
a random sampling method (such as Latin hypercube sampling),
since this is unlikely to yield a well-distributed test 
set given the small number of test cases we can produce.

\begin{figure}
	\centering
	\includegraphics[width=\columnwidth]{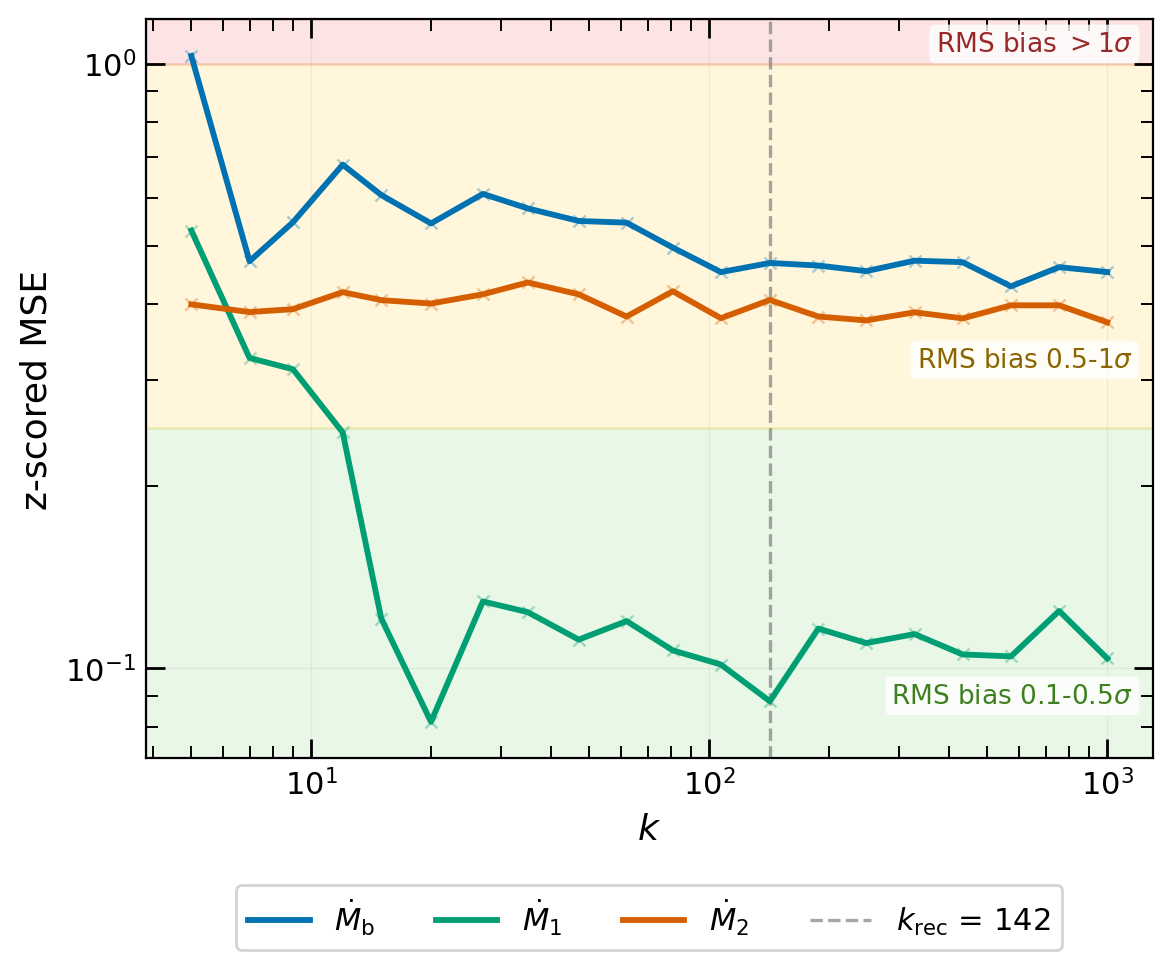}
	\caption{End-to-end zMSE of the full \calypso\ pipeline
	((\eb,\qb) interpolation followed by PCA reconstruction) on the test set,
	as a function of the number of PCA components $k$ retained,
	shown for each component of the time-series data:
	\Mbd\ in blue, $\dot{M}_1$ in orange, and $\dot{M}_2$ in green.
	The zMSE is computed by drawing synthetic realisations from the emulator at
	each test (\eb,\qb) and comparing the mean of these realisations
	to the mean of the held-out test windows.
	The colored shaded regions indicate the RMS bias 
	for a given range of zMSE values, with a RMS bias of 0 indicating a perfect emulator.
	A Pareto criterion across the three components is used to determine the 
	optimal number of PCA components to retain in the final model, 
	$k_{\rm final}$, which is indicated by the vertical dashed line and annotated marker.}
	\label{fig:k_sweep}
\end{figure}

First, we derive the number of PCA components required 
to maximize the performance of the emulator on the test set, 
while minimizing the dimensionality of the model ($k_{\rm final}$).
Figure \ref{fig:k_sweep} shows the z-scored mean squared error (zMSE) of \calypso\ 
predictions on the test set as a function of the number of PCA components retained in the model.
Specifically, the zMSE is computed as follows.
For each component of the time-series data, the residuals between 
the emulator predictions $\hat{y}$ and the true values $y$ are computed 
for each time point $i$ in the test set, yielding a vector of raw residuals,
\begin{equation}
	r_{(e_{\rm b},q_{\rm b}), i} = \hat{y}_i - y_i.
\end{equation}
The standard deviation of the true values across windows for each test case and each time bin
is then computed,
\begin{equation}
	\sigma_i = \sqrt{\frac{1}{N - 1} \sum_{j=1}^{N} (y_{j,i} - \bar{y}_i)^2},
\end{equation}
where $N$ is the number of windows for the given test binary and $\bar{y}_i$ is
the mean of the true values across those windows for time bin $i$.
The z-scored residuals are then computed by dividing the raw residuals 
by the standard deviation of the true values for each time bin,
\begin{equation}
z_{(e_{\rm b},q_{\rm b}), i} = \frac{r_{(e_{\rm b},q_{\rm b}), i}}{\sigma_i}.
\end{equation}
Finally, the zMSE for each test case is computed as the mean of the squared z-scored residuals over all time bins,
\begin{equation}
\text{zMSE}_{(e_{\rm b},q_{\rm b})} = \frac{1}{T} \sum_{i=1}^{T} z_{(e_{\rm b},q_{\rm b}), i}^2,
\end{equation}
and the reported zMSE is the median across all test cases,
\begin{equation}
\text{zMSE} = \mathrm{median} \left( \text{zMSE}_{(e_{\rm b},q_{\rm b})} \right).
\end{equation}

\Mbd, $\dot{M}_1$, and $\dot{M}_2$ are shown separately,
with \Mbd\ in blue, $\dot{M}_1$ in orange, and $\dot{M}_2$ in green
(a convention we use throughout the rest of the paper).
The MSE is computed as follows. 
For each test binary, we draw coefficient vectors from
the interpolator at truncation order $k$ and 
reconstruct $N_{\rm synth} = 500$ full time
series via the PCA basis.  We then compare the \emph{mean} of these
synthetic realisations to the mean of the held-out test windows.
This choice is deliberate: at a given $k$, each realisation is
reconstructed from a $k$-dimensional coefficient sample, and every
additional PCA component introduces an independent source of
stochastic variation in the reconstructed time series.  The total
reconstruction variance of a single realisation therefore grows with
$k$. 
Averaging over $N_{\rm synth}$ realizations suppresses the variance contribution,
isolating the bias and providing a cleaner diagnostic of how emulator
fidelity depends on the truncation order.

We then proceed to finding the optimal number of PCA components 
to retain in the final model, $k_{\rm final}$, across all three components. 
For each component, we find the global minimum of the zMSE
across the values of k tested, and store the 
corresponding number of PCA components at which the zMSE is within 10\% of the global minimum.
We then take the intersection of these sets
across the three components, and find the minimum number of 
PCA components in this intersection, which gives us $k_{\rm final}$.
If no intersection exists, we take the smallest acceptable $k$ for
each component individually and adopt the largest of the three as
$k_{\rm final}$, ensuring that even the most demanding component is
adequately resolved. Throughout the rest of the paper, 
we use $k_{\rm final}$ PCA components in the model.

\begin{figure}
	\centering
	\includegraphics[width=\columnwidth]{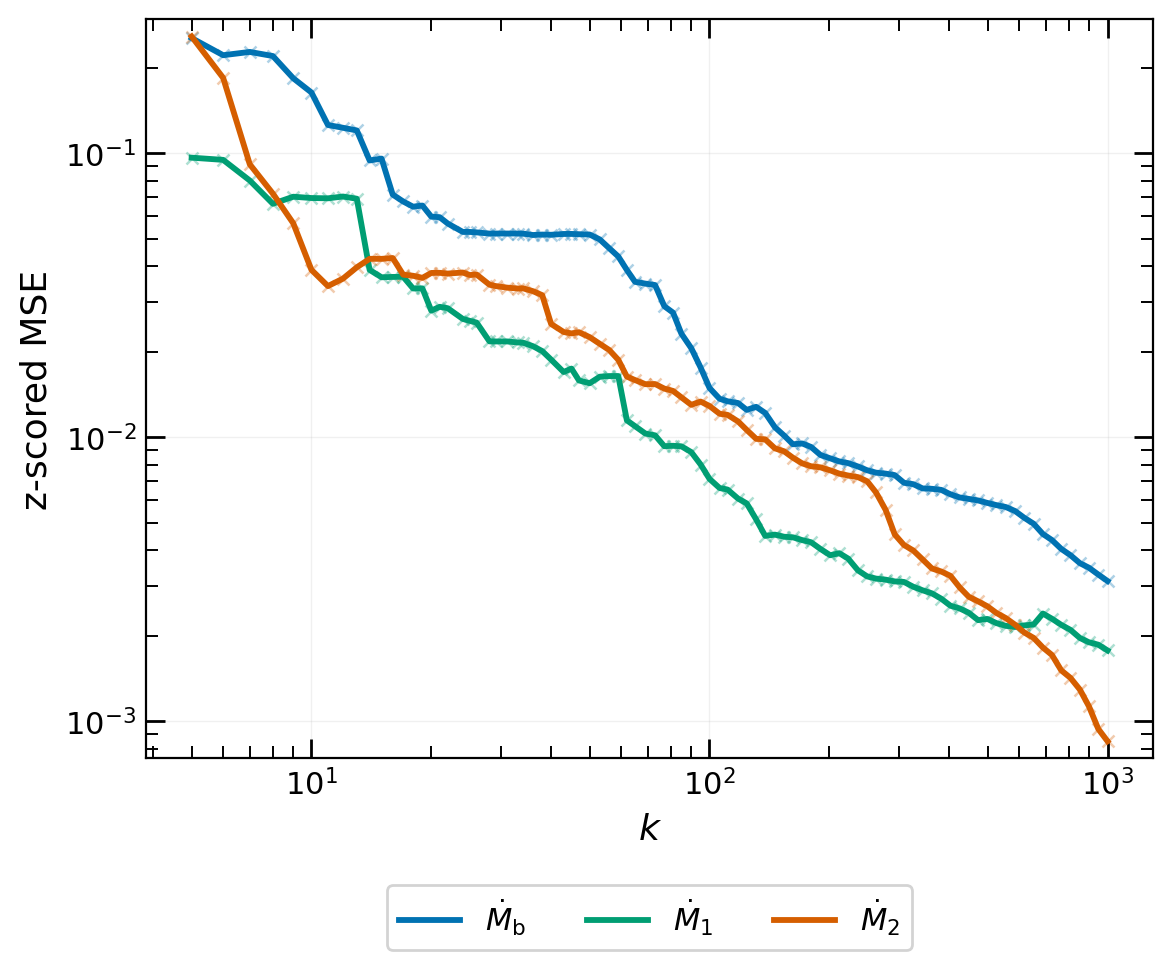}
	\caption{Basis-only zMSE on the test set as a function of the number of
	PCA components $k$, isolating the PCA basis from the (\eb,\qb) interpolator.
	Each test time series is projected directly onto the first $k$ basis vectors
	and reconstructed; the zMSE is then computed against the original time series,
	with no interpolation step involved.
	Comparison with Figure \ref{fig:k_sweep} therefore disentangles how much of
	the end-to-end error originates from basis truncation versus from the
	(\eb,\qb) interpolation.
	As in Figure \ref{fig:k_sweep}, $\dot{M}_1$ outperforms $\dot{M}_2$ and $\dot{M}_{\rm b}$,
	indicating that the PCA basis captures the variance in $\dot{M}_1$
	more efficiently than in \Mbd\ and $\dot{M}_2$.}
	\label{fig:test_capture}
\end{figure}

We note that the zMSE scores in Figure \ref{fig:k_sweep} differ significantly
across the three components, with $\dot{M}_1$ having the lowest zMSE and
$\dot{M}_2$ and \Mbd\ having higher, comparable zMSEs.
To diagnose where this per-component discrepancy originates, we isolate
the contribution of the PCA basis from that of the (\eb,\qb) interpolator.
In Figure \ref{fig:test_capture}, each test time series is projected directly
onto the first $k$ basis vectors and reconstructed, with no interpolation step
involved, and the zMSE is computed between the reconstructed and original
time series at each $k$. This `basis-only' diagnostic measures how efficiently
the PCA basis itself captures variance in each component, independent of the
emulator's parameter-space interpolation, and is shown alongside the
end-to-end zMSE of Figure \ref{fig:k_sweep} to disentangle the two error
contributions.
We find that the PCA basis captures the variance in $\dot{M}_1$ more efficiently
than in \Mbd\ and $\dot{M}_2$,
which is consistent with the lower zMSE of $\dot{M}_1$ in Figure \ref{fig:k_sweep}.
This indicates that the PCA basis itself is more efficient
at capturing the variations in time-series properties of $\dot{M}_1$
than \Mbd\ and $\dot{M}_2$, which is likely related to the fact
that $\dot{M}_1$ has lower variability amplitude and a more regular variability
pattern across the parameter space than \Mbd\ and $\dot{M}_2$.

\begin{figure*}
	\centering
	\includegraphics[width=\textwidth]{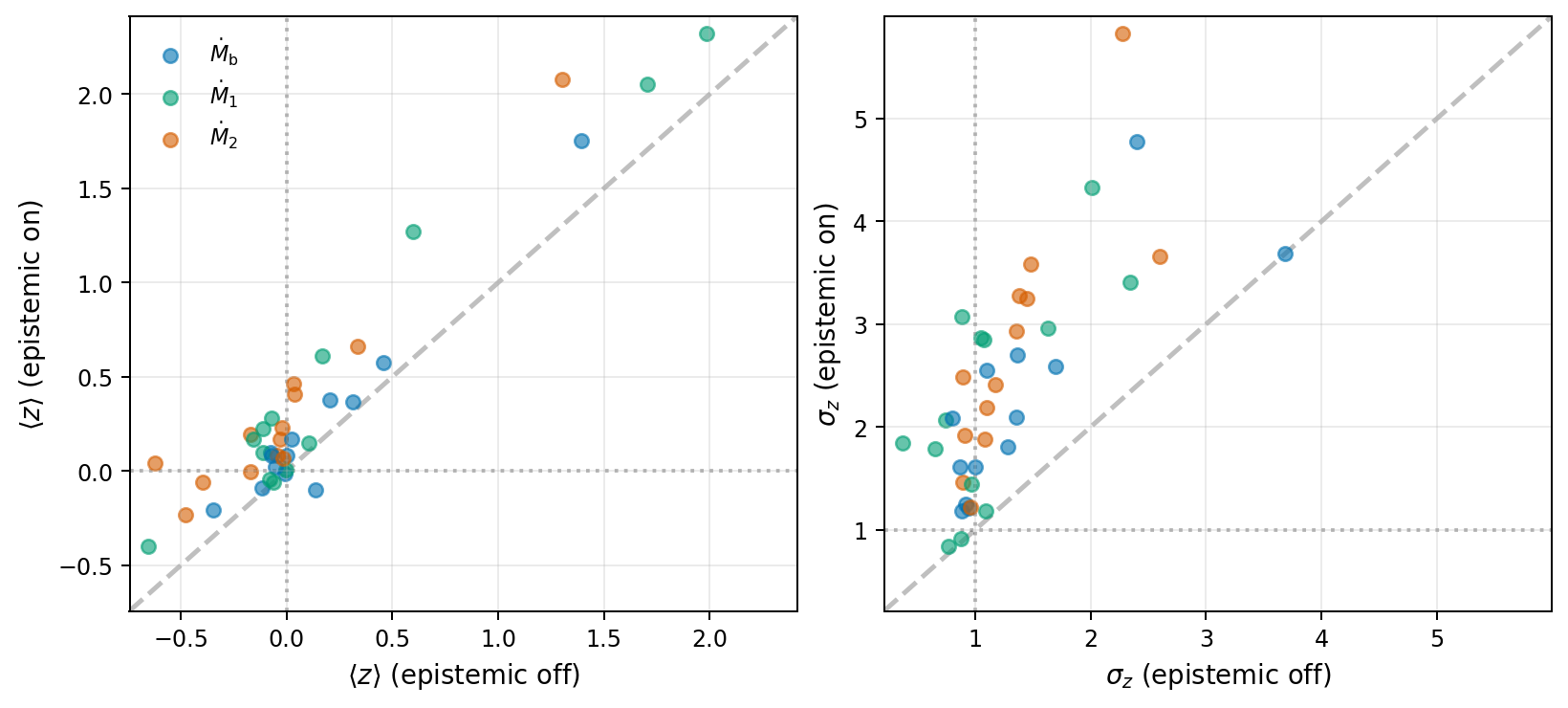}
	\caption{Effect of including the epistemic uncertainty on the
	calibration of \calypso\ predictions on the test set, plotted as
	scatter of the metric value with epistemic OFF (x-axis) versus the
	same metric with epistemic ON (y-axis); each point is one of the
	13 held-out test binaries, coloured by component
	($\dot M_{\rm b}$, $\dot M_1$, $\dot M_2$). The dashed black line
	marks $y=x$ (no change between settings); the dotted lines mark the
	perfectly-calibrated values.
	Left: mean of the per-time-bin z-score, $\langle z \rangle$,
	which captures the bias of the emulator predictions relative to the
	test set; the ideal value is $\langle z \rangle = 0$. The bias is
	systematically larger (in absolute value) when epistemic uncertainty
	is included.
	Right: standard deviation of the per-time-bin z-score,
	$\sigma_z$, which captures the dispersion of the predictions
	relative to the test scatter; the ideal value is $\sigma_z = 1$.
	With epistemic uncertainty on, $\sigma_z$ shifts above unity for all
	three components, indicating that the emulator draws from a
	distribution that is too wide compared to the test data.
	$\langle z \rangle$ and $\sigma_z$ together define the
	bias--dispersion decomposition of $z_{\rm RMS}$
	(equation~\ref{eq:z_rms}).}
	\label{fig:epistemic_comparison}
\end{figure*}

In Figure \ref{fig:epistemic_comparison}, 
we show the effect of including the epistemic uncertainty 
on the emulator performance on the test set.
For this plot, we choose an extended metric over the zMSE 
utilized in the previous Figures \ref{fig:k_sweep} and \ref{fig:test_capture}. 
Instead of the zMSE of the mean of the synthetic realisations compared to 
the mean of the test windows, 
which captures the absolute bias of the emulator, 
we compute the $z_{\rm RMS}$ error, 
which captures both bias and variance of the emulator predictions compared to the test set.
We first compute the z-score $z_i$ in each timebin of each window for each
test case,
\begin{equation}
z_i = \frac{\hat{y}_{\rm w, i} - y_i}{\sigma_{i}},
\end{equation}
where $\sigma_i$ is the standard deviation of the true values 
across windows for each test case and each time bin, 
$\hat{y}_{\rm w, i}$ is the value of the emulator prediction 
for a single synthetic realisation in time bin $i$, and 
$y_i$ is the mean of the true values of the held-out
test case across windows for time bin $i$.

The distribution of these z-scores across all timebins and windows for 
each test case is then summarized by the root mean square (RMS) of the z-scores,
\begin{equation}
z_{\rm RMS} = \sqrt{\langle z^2 \rangle} = \sqrt{ \langle z \rangle^2 + \sigma_z^2 },
\label{eq:z_rms}
\end{equation}
where $\langle z \rangle$ is the mean of the z-scores across 
all timebins and windows for a given test case, 
and $\sigma_z$ is the standard deviation of the z-scores 
across all timebins and windows for that test case.
Thus, $\langle z \rangle$ isolates the bias of the emulator predictions compared to the test set,
while $\sigma_z$ isolates the variance of the emulator predictions compared to the test set.
In a perfect emulator, we would expect the z-scores to be distributed as a standard normal distribution,
with $\langle z \rangle = 0$ and $\sigma_z = 1$, yielding $z_{\rm RMS} = 1$ as 
the ideal value.

In Figure \ref{fig:epistemic_comparison},
where we compare the performance of the emulator with and without epistemic uncertainty,
the realizations $\hat{y}_{\rm w, i}$ are drawn from the multivariate Gaussian distribution 
over the PCA coefficients with and without the epistemic variance term, respectively.
In the left panel of Figure \ref{fig:epistemic_comparison}, 
we show the $\langle z \rangle$ term of the $z_{\rm RMS}$ error,
which measures the bias of the emulator predictions compared to the test set.
We find that including the epistemic uncertainty causes the bias to increase for all three components,
indicating that drawing from a distribution with a larger variance leads to 
a worse match between the emulator
predictions and the test set, and therefore a larger bias.
The right hand side of Figure \ref{fig:epistemic_comparison} shows the $\sigma_z$ term of the $z_{\rm RMS}$ error,
which measures the variance of the emulator predictions compared to the test set.
We find that including the epistemic uncertainty causes the variance to increase for all three components.
Specifically, the variance
increase to values above 1 for all three components, 
indicating that the emulator predictions 
draw from a too-wide distribution compared to the test set.
We therefore do not include the epistemic uncertainty going forward, 
or in the default settings of the final public release of \calypso,
since it does not improve the performance of the emulator on the test set.

\section{Evaluation of Emulator Performance across the Parameter Space}
\label{sec:evaluation}
In the following section we evaluate the performance of \calypso\ across the (\eb,\qb) parameter space.
While the previous section focused on validating the performance of \calypso\ on the test set as a whole,
this section focuses on
investigating how the emulator performance varies across the parameter space,
and whether there are specific regions of the 
parameter space where the emulator performs better or worse.

\begin{figure*}
    \centering
    \includegraphics[width=0.85\textwidth]{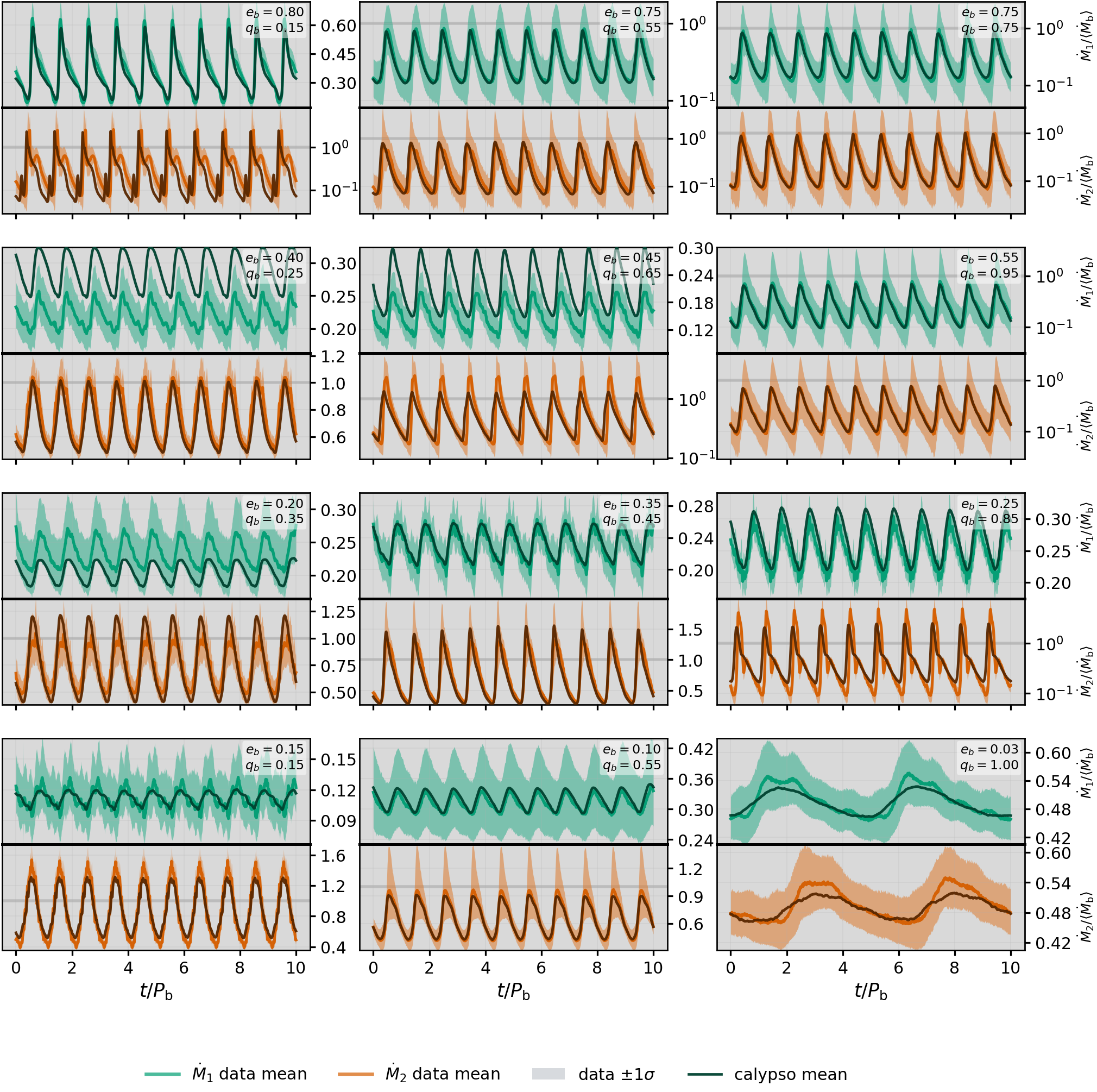}
    \caption{\calypso\ predictions versus held-out test
    simulations for 12 of the 13 test (\eb,\qb) pairs, arranged on
    the parameter-space grid (\eb\ increasing bottom to top, \qb\
    increasing left to right; cf.\ Figure~\ref{fig:pspace_test}).
    Each cell contains two stacked sub-panels: $\dot M_1$ (top, green)
    and $\dot M_2$ (bottom, orange). The lighter line and shaded band
    show the mean and $\pm 1\sigma$ scatter across 500 phase-aligned
    held-out windows; the darker line is the mean of
    $N_{\rm synth}=500$ stochastic \calypso\ realisations
    (epistemic off). The horizontal axis spans one 10-orbit window
    in units of $P_{\rm b}$; the vertical axis is
    $\dot M_i / \langle \dot M_{\rm b} \rangle $. 
	The largest biases occur in the primary variability of the binaries
    at (\eb=0.4, \qb=0.25) and (\eb=0.2, \qb=0.35).}
	\label{fig:test_timeseries}
\end{figure*}

In Figure \ref{fig:test_timeseries}, we show the time-series data for the test set,
with primary accretion rates on the top (green), and secondary accretion rates on the bottom (orange).
We again distribute the test cases across the parameter space
with eccentricity increasing from bottom to top and mass ratio increasing from left to right, as in Figure 
\ref{fig:pspace_test}.
The lighter green and orange lines and shaded regions 
show the mean and standard deviation of the test windows, respectively,
while the darker green and orange lines show the 
mean of the emulator predictions across 
$N_{\rm synth} = 500$ synthetic realisations.
We find that the emulator performs well in capturing the variability patterns in the test set,
noting that the most significant biases are in the primary variability of the 
binaries with \eb=0.4, \qb=0.25 and \eb=0.2, \qb=0.35.
However, in most other cases, the emulator predictions are well
within the standard deviation of the test windows.
The variability patterns of all binaries are well captured,
in particular the periodicity and amplitude of the variability.

\begin{figure*}
	\centering
	\includegraphics[width=0.85\textwidth]{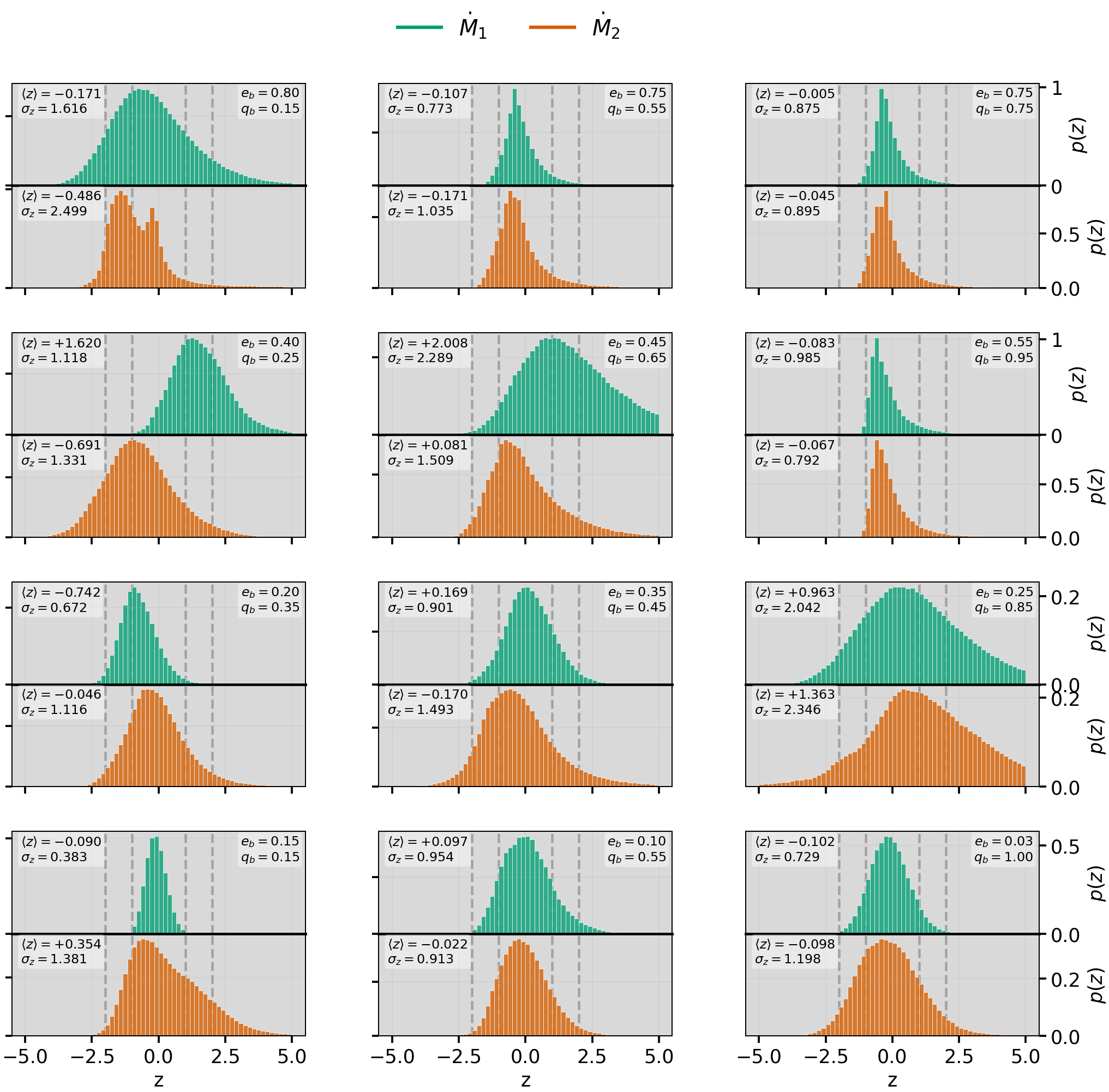}
	\caption{Per-binary z-score distributions for the same 12 of 13
	test (\eb,\qb) pairs as in Figure~\ref{fig:test_timeseries},
	with the same grid layout and the same component colours
	($\dot M_1$ top/green, $\dot M_2$ bottom/orange). The z-score
	is computed per time bin, by subtracting the mean of the held-out
	test windows from each synthetic emulator realisation and normalising
	by the standard deviation of the test windows.
	Each histogram aggregates $z$-scores over all time bins of
	$N_{\rm synth} = 500$ phase-aligned \calypso\ realisations and
	is density-normalised. Faint dashed verticals mark $z=\pm 1, \pm 2$.
	The upper-left text box reports the panel's $\langle z\rangle$
	and $\sigma_z$; the upper-right box repeats (\eb,\qb). Perfect
	calibration corresponds to $\langle z\rangle=0$, $\sigma_z=1$
	(see also Figure~\ref{fig:epistemic_comparison}).}
	\label{fig:zhist}
\end{figure*}

In Figure \ref{fig:zhist}, 
we show the distribution of the z-scores of the emulator predictions
compared to the test set for all time bins and windows. Specifically,
we aggregate the z-scores across all time bins and windows,
forming a distribution of z-scores for each test case and each component of the time-series data.
For each distribution, we compute the mean and standard deviation of the z-scores,
which are shown in a text box in the upper-left of each panel.
As in Figure \ref{fig:test_timeseries}, each test case and binary component
is shown in a separate panel, following the same color scheme as in Figure \ref{fig:test_timeseries}. 
The test cases are distributed across the parameter space
with eccentricity increasing from bottom to top and mass ratio increasing from left to right.
We further plot the 1 and 2 $\sigma$ intervals of a 
standard normal distribution as vertical dashed lines for reference.

\begin{figure*}
	\centering
	\includegraphics[width=\textwidth]{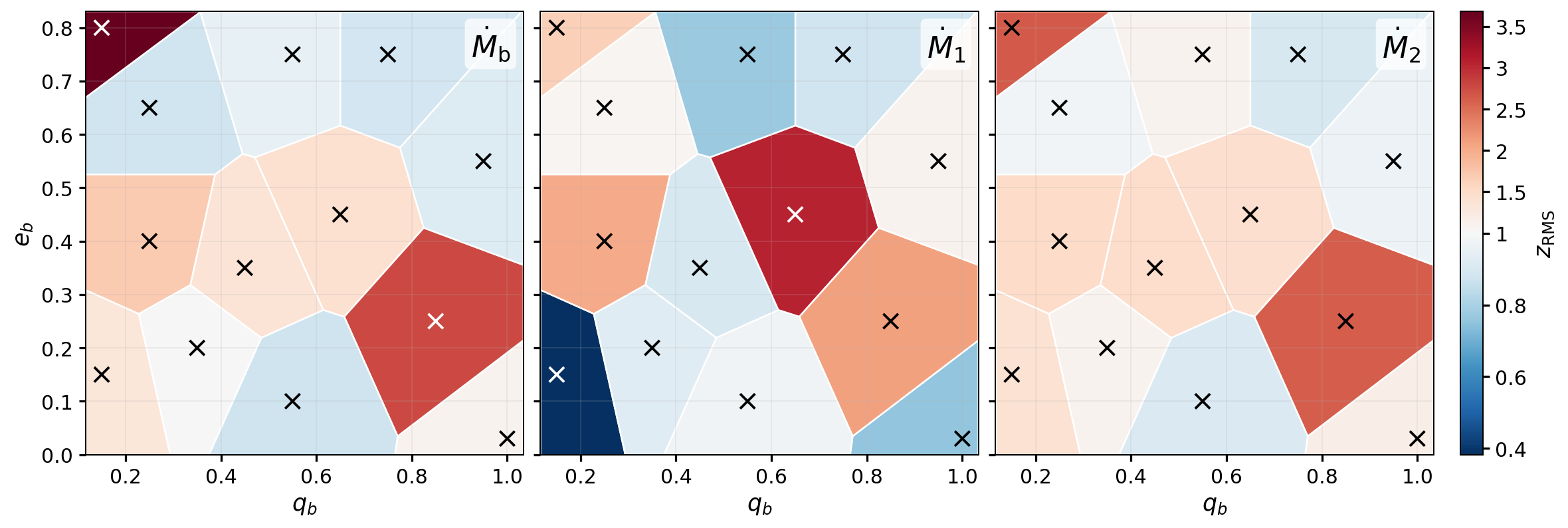}
	\caption{Per-binary $z_{\rm RMS}$ across the 13 held-out test cases
	in the (\eb,\qb) parameter space. From left to right, panels show
	the binary, primary, and secondary accretion-rate components
	($\dot M_{\rm b}$, $\dot M_1$, $\dot M_2$); the horizontal axis is
	\qb\ and the vertical axis is \eb. Each cross marks a test case, and
	the surrounding Voronoi cell is colored by that case's
	$z_{\rm RMS}$ (defined in \S\ref{sec:validation}), aggregated over
	all time bins of $N_{\rm synth}=500$ phase-aligned \calypso\
	realisations. The diverging RdBu colormap (shared colorbar, right)
	is anchored at $z_{\rm RMS}=1$, the value expected when the emulator
	predictive scatter matches the held-out scatter: red cells flag
	$z_{\rm RMS}>1$ (emulator biased or under-dispersed), blue cells
	flag $z_{\rm RMS}<1$ (over-dispersed).}
	\label{fig:test_voronoi}
\end{figure*}

In Figure \ref{fig:test_voronoi}, we investigate the performance of
the emulator across the (\eb,\qb) parameter space 
by plotting the $z_{\rm RMS}$ error of the emulator predictions for each test case.
Around the set of test cases in the 
parameter space, we construct a Voronoi tessellation 
and color each Voronoi cell by the $z_{\rm RMS}$ error of the corresponding test case.
As explained in \S\ref{sec:validation},
the $z_{\rm RMS}$ error captures both the bias and variance 
of the emulator predictions compared to the test set, 
with a value of 1 indicating a perfect emulator.
The choice of colormap is such that values close to 1 are shown in white, 
while values significantly above 1 are shown in red, 
indicating a worse match between the emulator predictions and the test set.
Blue values have lower $z_{\rm RMS}$ than 1, indicating 
that the variance of the emulator predictions is smaller than the variance of the test set.

We find that the binary with \eb=0.8 and \qb=0.15 has the highest $z_{\rm RMS}$, 
which is likely related to the fact that this binary has a high 
variability amplitude and a complex variability pattern across the parameter space, 
(see Figure \ref{fig:test_timeseries}),
making it more difficult for the PCA basis to capture its 
variability efficiently (see Figure \ref{fig:test_capture}).
This binary is also near the edge of the parameter space covered by the training data,
which may contribute to the higher $z_{\rm RMS}$ due to the interpolation step of
the emulator.

\section{Application: Synthetic LSST Light Curves}
\label{sec:lsst_application}

A central motivation for \calypso\ is the rapid generation of synthetic
accretion light curves anywhere in the (\eb,\qb) parameter space, which
can then be combined with a radiative model to produce observable signatures.
In this section we illustrate the end-to-end pipeline by converting
\calypso-emulated accretion-rate samples into apparent magnitudes in the
LSST $g$-band, for a set of binary configurations chosen to lie
\emph{off} the training and test grids.
A reference implementation of this pipeline is included in the public
\calypso\ release.

\subsection{From accretion rate to apparent magnitude}
\label{sec:lum_pipeline}

For a binary with total mass \Mb, mass ratio \qb $\equiv M_2/M_1 \in (0, 1]$,
and Eddington fraction $f_{\rm Edd}$, the physical accretion rate
onto each component $i \in \{1, 2\}$ is obtained by rescaling the
dimensionless \calypso\ outputs $\dot{m}_i \equiv \dot{M}_i / \dot{M}_{\rm Edd}$
by the binary-wide Eddington rate,
\begin{equation}
	\dot{M}_{\rm Edd} \;=\; \frac{f_{\rm Edd}\, L_{\rm Edd}(M_{\rm b})}{\eta\, c^2},
	\qquad
	L_{\rm Edd}(M_{\rm b}) \;=\; \frac{4\pi G M_{\rm b} m_p c}{\sigma_T},
	\label{eq:mdot_edd}
\end{equation}
where $\eta = 0.1$ is the radiative efficiency, $m_p$ the proton mass and
$\sigma_T$ the Thomson cross-section. The component mass enters the
mini-disk radiative model and is given by
$M_1 = M_{\rm b}/(1+q_{\rm b})$ and $M_2 = q_{\rm b}\,M_{\rm b}/(1+q_{\rm b})$.

Each accreting component is modelled as a geometrically thin,
optically thick Shakura–Sunyaev disk extending between an inner radius
$r_{\rm min} = 20\, G M_i / c^2$ (chosen to exclude the innermost region
that emits primarily in X-rays and is poorly described by the
thin-disk approximation) and an outer radius $r_{\rm max}$.
For simplicity in this section we adopt a steady-state circumbinary
prescription for the outer radius of each mini-disk, 
which is a fixed fraction of the binary separation \ab\, 
\begin{equation}
	r_{\rm max} \;=\; 0.27\, a_{\rm b},
	\label{eq:rmax_steady}
\end{equation}
which is independent of orbital phase and binary parameters other than \ab.
The local effective temperature of the disk follows the standard
thin-disk profile,
\begin{equation}
	T(R) \;=\; \left(\frac{3\, G M_i\, \dot{M}_i}{8 \pi \sigma R^3}\right)^{1/4},
	\label{eq:T_disk}
\end{equation}
where $\sigma$ is the Stefan–Boltzmann constant.
Each annulus radiates as a blackbody, so that the spectral luminosity
of component $i$ at rest-frame wavelength $\lambda_{\rm em}$ is
\begin{equation}
	L_{\lambda, i}(\lambda_{\rm em}, t) \;=\;
	\int_{r_{\rm min}}^{r_{\rm max}} 4\pi^2 R\, B_\lambda\!\left(\lambda_{\rm em}, T(R, t)\right)\, dR,
	\label{eq:Llambda}
\end{equation}
with $B_\lambda$ the Planck function.
The full binary spectrum is the sum
$L_\lambda(t) = L_{\lambda,1}(t) + L_{\lambda,2}(t)$.

To translate $L_\lambda$ into an LSST-band AB magnitude at redshift $z$,
we integrate over the rest-frame band that maps to the observed bandpass,
$[\lambda^{\rm em}_{\min}, \lambda^{\rm em}_{\max}] = [\lambda^{\rm obs}_{\min}, \lambda^{\rm obs}_{\max}] / (1+z)$.
A $K$-correction relates the resulting band-integrated rest-frame
luminosity $L_{\rm band}$ to the observed monochromatic flux density
at the bandpass pivot wavelength $\lambda_{\rm pivot}^{\rm obs}$
\citep[e.g.,][]{Hogg2002_K},
\begin{equation}
	f_\nu \;=\; \frac{L_{\rm band}\,\left(\lambda_{\rm pivot}^{\rm obs}\right)^2}
	                {4\pi\, D_L^2(z)\, (1+z)\, \Delta\lambda_{\rm em}\, c},
	\label{eq:Kcorr}
\end{equation}
where $D_L(z)$ is the luminosity distance for a flat $\Lambda$CDM
cosmology ($H_0 = 70$~km~s$^{-1}$~Mpc$^{-1}$, $\Omega_m = 0.3$) and
$\Delta\lambda_{\rm em}$ is the rest-frame bandpass width.
The apparent AB magnitude is then
\begin{equation}
	m_{\rm AB} \;=\; -2.5\, \log_{10}\!\left(\frac{f_\nu}{\mathrm{erg\,s^{-1}\,cm^{-2}\,Hz^{-1}}}\right) - 48.6.
	\label{eq:mAB}
\end{equation}

\subsection{On the choice of outer disk radius}
\label{sec:roche_note}

The steady-state assumption of equation~(\ref{eq:rmax_steady}) is a
deliberate simplification: in eccentric or unequal-mass binaries
the per-component Roche lobe contracts and expands on the orbital
timescale. This effect can be seen by eye in hydrodynamic simulations,
and adds a time-varying component to the disk structure.
A more physical prescription truncates each mini-disk at
its instantaneous Roche radius. Using the Eggleton approximation
\citep{Eggleton1983},
\begin{equation}
	r_{\rm ER}(t) \;=\; r_{\rm b}(t) \cdot
	\frac{0.49\, q^{2/3}}{0.6\, q^{2/3} + \ln\!\left(1 + q^{1/3}\right)},
	\label{eq:ER}
\end{equation}
with the instantaneous binary separation
$r_{\rm b}(t) = a_{\rm b}\,[1 - e_{\rm b}\,\cos E(t)]$ and $E(t)$ the
eccentric anomaly, the outer radius modulates with the orbital phase
and depends on the mass ratio of the considered component to its companion
($q = M_i / M_{j\neq i}$).

This time-varying outer radius introduces an additional, purely geometric
source of variability that can amplify the optical-band variability
peaks emerging from the underlying accretion-rate oscillations
captured by \calypso. The amplitude of this effect is non-trivial and depends sensitively
on the assumed disk thermal response time relative to the orbital period.
A self-consistent treatment requires coupling the disk thermal/viscous
evolution to the orbital phase, which is beyond the scope of this paper.
We therefore adopt the simpler steady prescription throughout this section,
and note that $r_{\rm ER}(t)$ should be considered when interpreting
periastron-phased modulations in observed candidates.

\subsection{Synthetic LSST $g$-band light curves}
\label{sec:lc_showcase}

\begin{figure*}
	\centering
	\includegraphics[width=\textwidth]{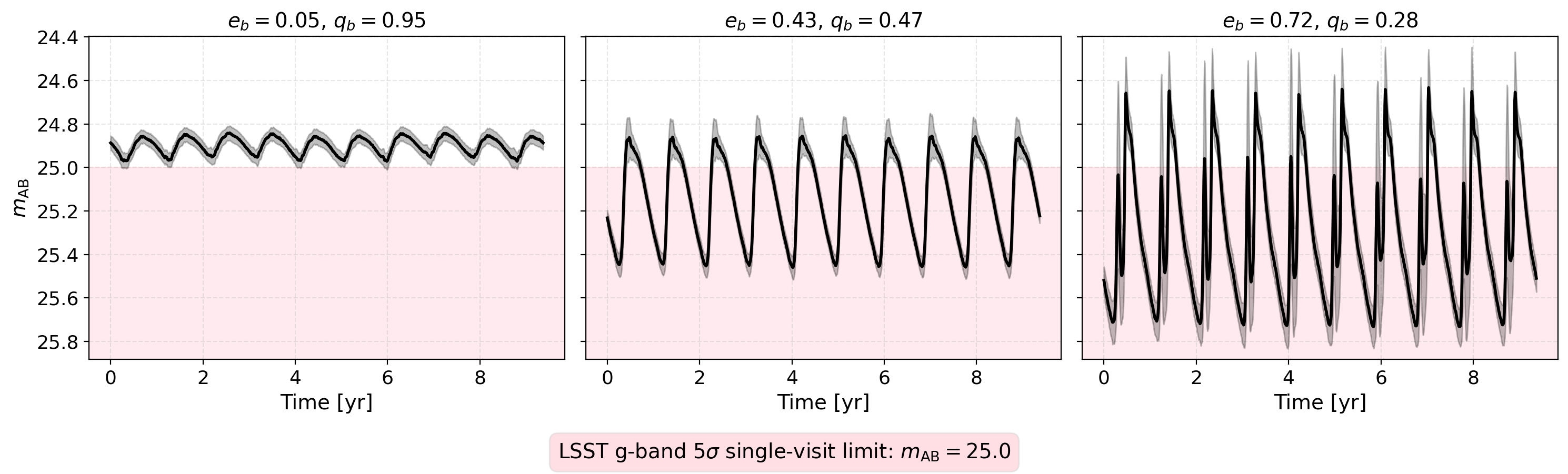}
	\caption{Synthetic LSST $g$-band apparent magnitudes for three
	off-grid \calypso\ binary configurations, showing the binary-total
	signal $\dot{M}_{\rm b}$.
	Mean and 1$\sigma$ shaded bands are computed across $N_{\rm draws} = 100$
	stochastic \calypso\ samples per case.
	Reference parameters: total binary mass
	$M_{\rm b} = 10^7\,\msol$, semi-major axis
	$a_{\rm b} = 10^{-3}\,\mathrm{pc}$ corresponding to an orbital period
	$P_{\rm b} \approx 0.94\,\mathrm{yr}$, redshift $z = 1.0$,
	Eddington fraction $f_{\rm Edd} = 0.1$, and radiative efficiency
	$\eta = 0.1$.
	The mini-disk extends from $r_{\rm min} = 20\, GM_i/c^2$ to
	$r_{\rm max} = 0.27\, a_{\rm b}$ as in
	equation~(\ref{eq:rmax_steady}).
	The pink shaded band marks the LSST $g$-band $5\sigma$ single-visit
	depth ($m_{\rm AB} = 25.0$); fluctuations dipping into this band fall
	below per-visit detectability but remain accessible in coadded stacks.
	The (\eb,\qb) cases shown — $(0.05, 0.95)$, $(0.43, 0.47)$,
	$(0.72, 0.28)$ — were chosen to lie strictly between training-grid
	points to highlight that the curves are fully synthesized via
	Cholesky-space interpolation in the (\eb,\qb) plane.}
	\label{fig:lum_lightcurves}
\end{figure*}

Figure~\ref{fig:lum_lightcurves} shows the resulting binary apparent
magnitudes for three off-grid binary configurations chosen to span
distinct dynamical regimes: a near-circular equal-mass binary,
a moderately eccentric and moderately unequal-mass system,
and a strongly eccentric, asymmetric system.
None of the (\eb,\qb) values lie on the \calypso\ training or test
grids — they are fully synthesized via Cholesky-space interpolation.
The mean and 1$\sigma$ scatter are computed across 100 stochastic
\calypso\ draws per case.
For the chosen reference configuration ($M_{\rm b} = 10^7\,M_\odot$,
$z = 1$, $f_{\rm Edd} = 0.1$), all three regimes sit within $\sim 1$~mag
of the LSST $g$-band $5\sigma$ single-visit limit, with the
eccentric configurations dipping below the limit at periastron.
We contrast three such cases to show that, for a given set of binary parameters, 
the variability amplitude and pattern can differ significantly, 
with the more eccentric and unequal-mass systems showing 
stronger variability and sharper peaks at periastron 
than the near-circular equal-mass system. 
The binary parameters will likely
have a significant effect on the variability pattern, 
and therefore on the detectability of the binary in time-domain surveys, 
in particular when the apparent magnitude is close to the survey detection limit.

\section{Summary and Discussion}
\label{sec:discussion}
We have developed \calypso, a fully stochastic PCA-based emulator for the
accretion variability of eccentric, unequal-mass binaries across
a wide range of eccentricities and mass ratios.
Trained on a suite of 2D hydrodynamic simulations of circumbinary accretion,
\calypso\ reproduces the time-series data on which it was trained and
interpolates smoothly in the (\eb,\qb) parameter space
using a multivariate Gaussian distribution over the PCA coefficients.
A z-scored MSE estimate on the test set was used to select the
number of PCA components retained in the final model, yielding
$k_{\rm final} = 142$ components and capturing $\gtrsim 90\%$ of the variance
in the training data. Beyond $k_{\rm final}$, the zMSE on the test set plateaus,
so additional components are omitted to keep the final model compact.

We have validated \calypso\ on a test set of 13 (\eb,\qb) pairs that were
not included in training, and find that it performs well in capturing the
variability patterns of the held-out simulations. In particular, the
periodicity and amplitude of the variability are well reproduced.
We do not identify any specific region of the parameter space where the
emulator performs significantly better or worse than another, with only a
few outliers showing larger biases (e.g.\ \eb=0.40, \qb=0.25 and
\eb=0.20, \qb=0.35; see Figure~\ref{fig:test_timeseries}) or higher variance
(e.g.\ \eb=0.25, \qb=0.85 and \eb=0.45, \qb=0.65; see Figure~\ref{fig:zhist})
than the rest of the test set.

A key feature of \calypso\ is its ability to capture both the short-term
stochastic variability in the accretion time series and the long-term
modulation due to disk precession. This is achieved by including multiple
time windows for each (\eb,\qb) pair in the training dataset, which allows
the PCA basis to capture the stochasticity of the accretion variability,
and by modelling the distribution of the PCA coefficients across these
windows as a multivariate Gaussian. This approach captures the aleatoric
uncertainty arising from the intrinsic variability of the accretion process.
We have also developed an estimate for the epistemic uncertainty, which
arises from the limited number of training samples and the interpolation
in the (\eb,\qb) parameter space. However, we find that including the
epistemic uncertainty does not improve performance on the test set; instead
it inflates the variance of the emulator predictions, and we therefore do
not include it in the default settings of \calypso. 
In other words, the aleatoric component alone already reproduces
the variance observed in the held-out simulations, and our particular
estimator of the epistemic uncertainty over-counts the additional spread
introduced by interpolation in the (\eb,\qb) plane beyond what the
validation data support.

We have further demonstrated the end-to-end pipeline of using \calypso\ to
generate synthetic accretion-rate time series for arbitrary (\eb,\qb) pairs
and to convert them into synthetic LSST $g$-band light curves with a simple
radiative model. For a binary with a given mass, separation, redshift and accretion rate, 
the variability pattern and amplitude
vary significantly across (\eb,\qb) pairs, with potentially significant
consequences for the detectability of such binaries in time-domain surveys.

\subsection{Caveats and Future Work}

A number of caveats apply to the present version of \calypso, both on the
training-data side and on the modelling assumptions in the emulator
and its downstream applications.

\calypso\ is trained on a specific suite of 2D hydrodynamic simulations,
and its predictions are only as good as the underlying simulations.
The training simulations adopt a locally isothermal equation of state,
a fixed disk aspect ratio, and a fixed Shakura-Sunyaev viscosity parameter
$\alpha$. These assumptions need not hold in real circumbinary disks,
whose thermodynamics can be more complicated; \calypso\ predictions
should therefore be interpreted accordingly when applied to observed
systems. \calypso\ also interpolates within the trained
(\eb,\qb) box but does not extrapolate beyond it, and the multivariate
Gaussian assumption on the PCA coefficients (validated in
Appendix \ref{sec:appendix_pca_gauss}) is an approximation that we have
verified empirically rather than derived from first principles.
Each emulator draw covers a fixed window of 10 binary orbits;
longer time series can be stitched from independent windows, but secular
drift on timescales $\gg P_{\rm b}$ is not captured by such stitching.
Finally, the radiative pipeline used in \S\ref{sec:lsst_application}
is itself idealized. It assumes a thin Shakura-Sunyaev disk with a steady-state
$r_{\rm max}=0.27\,a_{\rm b}$, no radiation reprocessing, wind or jet contribution,
or relativistic effects. Therefore, the synthetic LSST magnitudes
should be read as a transfer-function illustration of the emulator's
output rather than a dogmatic prediction.
On the diagnostics side, our primary performance metric is the per-time-bin z-score 
and its summary statistics (Figures \ref{fig:test_capture}, \ref{fig:zhist}).
The z-scored MSE uses only the marginal standard deviation at each time bin 
and therefore captures errors in the marginal predictive distribution, 
but is insensitive to errors in the temporal covariance of the emulated time series.
It does not capture how well \calypso\ reproduces correlations 
between neighbouring (or distant) time bins. We discuss potential
future diagnostics that explicitly probe the temporal covariance below.

Several extensions are natural follow-ups to this work.
First, the emulator can be augmented with additional binary and disk
parameters such as the disk aspect ratio and viscosity $\alpha$, as more
simulations covering these dimensions of the parameter space become
available.
Second, \calypso\ can be extended to emit time series of user-specified
length, rather than the fixed 10-orbit window of the current version. For
now, users can stitch together independent 10-orbit predictions to obtain
longer time series, with the secular caveat noted above.
Third, we plan to disentangle the short-term stochastic variability from
the long-term modulation due to disk precession. One route is to add the
precession angle as a training-time parameter: variability at fixed
precession angle could then be attributed to the stochastic component,
while variability across precession angles would isolate the secular
modulation.
Fourth, we plan to perform parameter inference on the held-out
simulations using the closed-form likelihood derived in
Appendix~\ref{sec:appendix_likelihood}, to quantify how tightly
(\eb,\qb) can be recovered from a noisy variability pattern,
which is the natural prerequisite to applying \calypso\ to real
time-domain candidates from surveys such as LSST\@.
We anticipate using \calypso\ in tandem with public radiative-transfer
tools such as \texttt{binlite} \citep{DOrazio2024}, which is currently
being extended to unequal-mass binaries, to translate emulated
accretion-rate time series into synthetic photometry and spectra.
Fifth, in addition to the per-time-bin z-score used here, 
future evaluations of the emulator can use statistics that 
explicitly probe the temporal covariance in the time series.
Examples include auto- and cross-correlation functions of the accretion rates, 
or the Mahalanobis distance under a non-diagonal time-bin covariance estimate.

To our knowledge, \calypso\ provides the first stochastic emulator of
accretion variability across the eccentric, unequal-mass binary
parameter space. As 3D hydrodynamic and MHD simulations, broader
disk-parameter coverage, and coupled radiative pipelines become
available, they can be folded into \calypso\ without restructuring its
PCA and interpolation backbone, so that the emulator can continue to
serve as a bridge between hydrodynamic predictions and the time-domain
surveys that motivated it.

\section{Installation and Usage}
\label{sec:installation}
We have developed \calypso\ as a Python package,
and it can be downloaded and installed via pip from the 
Python Package Index (PyPI) as follows,
\begin{verbatim}
pip install calypso-emulator
\end{verbatim}
with the main emulator usage pattern,
\begin{verbatim}
import calypso
emulator = calypso.load_emulator()
synth = emulator.predict(eb, qb)
\end{verbatim}
where \texttt{eb} and \texttt{qb} are the eccentricity 
and mass ratio of the binary, respectively,
and \texttt{synth} contains the synthetic time-series data 
generated by the emulator for the specified parameters.

\section{Acknowledgements}
\label{sec:ack}
This work was supported by a grant from the 
Simons Foundation International [SFI-MPS-SFJ-00006123, MSS].
This research was supported in part by grant
NSF PHY-1748958 to the Kavli Institute for Theoretical Physics (KITP).
MS acknowledges early conversations about this project with 
Maria Charisi at the KITP Circumbinary Disk workshop (2022), 
Steve Taylor and Lars Hernquist, and 
later discussions with Zoltan Haiman and Andrew MacFadyen.
MS thanks Chiara Mingarelli for helpful suggestions on the 
public-facing \calypso\ demo.
M.H. is supported by the Simons Collaboration on ``Learning the Universe''.

The computations in this paper were run on the FASRC Cannon cluster
supported by the FAS Division of Science Research Computing Group at
Harvard University.
The hydrodynamic simulations were performed with the \texttt{Arepo}
code \citep{Springel2010, Pakmor2016}.
This work made use of the following open-source software:
\texttt{numpy} \citep{Harris2020}, \texttt{scipy} \citep{Virtanen2020}, 
\texttt{matplotlib} \citep{Hunter2007}, \texttt{scikit-learn} 
\citep{Pedregosa2011}, and \texttt{astropy} \citep{Astropy2022}.

\section*{Data Availability}
The \calypso\ source code is publicly available on Zenodo
\citep{calypso_zenodo} and developed openly on GitHub at
\url{https://github.com/mssiwek/calypso}.
The GitHub repository includes documentation on how to use \calypso\ to make predictions for
arbitrary (\eb,\qb) pairs within the parameter space covered by the training data,
as well as example scripts and Jupyter notebooks to demonstrate its usage.
The PCA training basis,
interpolation tables, and hydrodynamic time series
underlying this work are archived on Zenodo
\citep{calypso_data_zenodo}. An interactive web demo is available at
\url{https://calypso.streamlit.app/}.

\appendix 

\section{Likelihood for parameter inference}
\label{sec:appendix_likelihood}

A natural application of \calypso\ is inference of
the binary parameters (\eb, \qb) from observed accretion-rate time series.
Our generative model is a multivariate Gaussian over
a finite set of PCA coefficients
(\S\ref{sec:training}, \S\ref{sec:interpolation}), so the corresponding
likelihood is also Gaussian and we can write it down in closed form.

Let $y_{\rm obs} \in \mathbb{R}^{3T}$ be an observed accretion time
series, formed by concatenating the de-trended, time-aligned
$\dot M_{\rm b}$, $\dot M_1$, and $\dot M_2$ identically to the
training pre-processing (\S\ref{sec:data_preprocessing}). Let
$V \in \mathbb{R}^{3T \times k}$ be the retained PCA basis matrix at
$k = k_{\rm final}$ with $V^\top V = I_k$, and
$\bar y \in \mathbb{R}^{3T}$ the training-set column mean. We assume
that $y_{\rm obs}$ is well represented by the first $k$ PCA components,
so that measurement noise and truncation residuals are small relative
to the in-basis variability of the model. The \calypso\ forward model
is then
\begin{equation}
y = \bar y + V c,
\qquad
c \sim \mathcal{N}\!\big(\boldsymbol{\mu}(e_{\rm b}, q_{\rm b}),\, \Sigma(e_{\rm b}, q_{\rm b})\big),
\label{eq:appendix_lik_forward}
\end{equation}
where $\boldsymbol{\mu}(e_{\rm b}, q_{\rm b})$ and
$\Sigma(e_{\rm b}, q_{\rm b})$ are the interpolated coefficient mean
and covariance from \S\ref{sec:interpolation}.

We first project the observation onto the basis. Because $V$ is
orthonormal, multiplying both sides of
equation~(\ref{eq:appendix_lik_forward}) by $V^\top$ inverts the basis
expansion and recovers the coefficient vector that corresponds to
$y_{\rm obs}$,
\begin{equation}
c_{\rm obs} = V^\top \big(y_{\rm obs} - \bar y\big) \in \mathbb{R}^k.
\label{eq:appendix_lik_proj}
\end{equation}
The projection $c_{\rm obs}$ is a deterministic, parameter-independent
function of $y_{\rm obs}$, so it carries the same information about
(\eb, \qb) as the original observation.

The forward model in equation~(\ref{eq:appendix_lik_forward}) gives us
a probability density over coefficients,
\begin{equation}
p\!\big(c \mid e_{\rm b}, q_{\rm b}\big) =
\frac{1}{\sqrt{(2\pi)^k\, |\Sigma|}}
\exp\!\Big[-\tfrac{1}{2} (c - \boldsymbol{\mu})^\top \Sigma^{-1} (c - \boldsymbol{\mu})\Big],
\label{eq:appendix_lik_density}
\end{equation}
with $\boldsymbol{\mu} = \boldsymbol{\mu}(e_{\rm b}, q_{\rm b})$ and
$\Sigma = \Sigma(e_{\rm b}, q_{\rm b})$ the interpolated mean and
covariance. Read as a function of $c$ at fixed parameters,
equation~(\ref{eq:appendix_lik_density}) is the standard normalised
density that integrates to one over $c \in \mathbb{R}^k$. We want to
read it instead as a function of the parameters at fixed observed
data, which is the definition of a likelihood. Specifically, we
define the likelihood as the probability density of the observed
$y_{\rm obs}$, taken as a function of the parameters with the data
held fixed,
\begin{equation}
L(e_{\rm b}, q_{\rm b} \mid y_{\rm obs}) \;\equiv\;
p\!\big(y_{\rm obs} \mid e_{\rm b}, q_{\rm b}\big)
\;=\; p\!\big(c_{\rm obs} \mid e_{\rm b}, q_{\rm b}\big) \cdot J,
\label{eq:appendix_lik_definition}
\end{equation}
where $J$ is the Jacobian factor from the change-of-variables formula,
which we now determine.

The change-of-variables formula for probability densities relates the
densities of two variables connected by a smooth invertible
transformation: the two densities differ by a multiplicative factor
that accounts for how the transformation locally stretches or
compresses volumes. For our linear projection
$c = V^\top(y - \bar y)$, that factor is
\begin{equation}
J \;=\; \frac{1}{\sqrt{\det\!\big(V^\top V\big)}},
\label{eq:appendix_lik_jacobian}
\end{equation}
which measures by how much the projection scales volumes on the
$k$-dimensional coefficient subspace. The columns of $V$ are PCA basis
vectors obtained from the SVD in \S\ref{sec:training} and are
unit-norm and pairwise orthogonal by construction, so $V^\top V = I_k$
and $\det(I_k) = 1$. The Jacobian factor is therefore $J = 1$:
geometrically, an orthonormal projection preserves lengths, and so it
preserves volumes; no volume element is stretched or compressed in
moving from $y_{\rm obs}$ to $c_{\rm obs}$. Equation~(\ref{eq:appendix_lik_definition})
then reduces to
\begin{equation}
L(e_{\rm b}, q_{\rm b} \mid y_{\rm obs}) =
p\!\big(c_{\rm obs} \mid e_{\rm b}, q_{\rm b}\big).
\label{eq:appendix_lik_definition_simplified}
\end{equation}

Taking the logarithm of equation~(\ref{eq:appendix_lik_density}) at
$c = c_{\rm obs}$ and collecting the parameter-independent
$-\tfrac{k}{2}\log(2\pi)$ Gaussian normalisation into a single
additive offset, we obtain
\begin{equation}
\begin{split}
\log L(e_{\rm b}, q_{\rm b} \mid y_{\rm obs}) ={}&
- \tfrac{1}{2}(c_{\rm obs} - \boldsymbol{\mu})^\top \Sigma^{-1} (c_{\rm obs} - \boldsymbol{\mu}) \\
&- \tfrac{1}{2}\log|\Sigma| + \mathrm{const}.
\end{split}
\label{eq:appendix_lik_coeff}
\end{equation}
The first term is the squared Mahalanobis distance between the
observed coefficient vector and the predicted mean, with deviations
weighted by the inverse covariance: it is small when $c_{\rm obs}$
lines up with $\boldsymbol{\mu}(e_{\rm b}, q_{\rm b})$, and large when
it does not. The
second term penalises parameter points where $\Sigma$ has a large
determinant. Without it, the Mahalanobis term alone would always
prefer broader covariances, since a wider $\Sigma$ shrinks
$\Sigma^{-1}$ and therefore reduces
$(c_{\rm obs} - \boldsymbol{\mu})^\top \Sigma^{-1} (c_{\rm obs} - \boldsymbol{\mu})$
regardless of how well $\Sigma$ actually matches the spread of the
residuals. The constant term collects
everything that does not depend on $(e_{\rm b}, q_{\rm b})$.
The cost of evaluating equation~(\ref{eq:appendix_lik_coeff}) at a
single $(e_{\rm b}, q_{\rm b})$ is dominated by the Cholesky-factor
interpolation of \S\ref{sec:interpolation}.

To turn the likelihood into a probability distribution over the
parameters, we apply Bayes' theorem with a prior
$p(e_{\rm b}, q_{\rm b})$ on the training-grid support,
\begin{equation}
\begin{split}
p(e_{\rm b}, q_{\rm b} \mid y_{\rm obs})
&= \frac{L(e_{\rm b}, q_{\rm b} \mid y_{\rm obs})\, p(e_{\rm b}, q_{\rm b})}{p(y_{\rm obs})} \\
&\propto L(e_{\rm b}, q_{\rm b} \mid y_{\rm obs})\, p(e_{\rm b}, q_{\rm b}).
\end{split}
\end{equation}
The denominator $p(y_{\rm obs})$ is the marginal likelihood, which does
not depend on $(e_{\rm b}, q_{\rm b})$ and is the normalisation that
makes the posterior integrate to one over the parameter space.
Standard MCMC or nested-sampling routines do not require it
explicitly, and so we can sample the posterior directly from the
unnormalised right-hand side.

Two caveats apply. First, the analytic likelihood assumes that
$y_{\rm obs}$ has been pre-processed identically to the training data,
with the same time grid, de-trending, and column-mean subtraction.
Real observations of accretion variability are typically light curves
rather than direct accretion rates, so applying
equation~(\ref{eq:appendix_lik_coeff}) requires recovering an estimate
of the accretion-rate time series $y$ from the observed magnitudes
$m$ before projecting onto the PCA basis. This means inverting the
radiative-transfer pipeline (\S\ref{sec:lsst_application})
 to get $y_{\rm obs}$
from $m$, and then applying
equations~(\ref{eq:appendix_lik_proj}) and~(\ref{eq:appendix_lik_coeff})
as written. The inversion is not always well-posed, since $m$ depends
on integrated disk quantities and may not be uniquely invertible. Second, the
likelihood inherits the modelling assumptions of the training
simulations (locally isothermal EOS, fixed $h/r$, fixed $\alpha$;
\S\ref{sec:hydro}), so posteriors on (\eb, \qb) should be interpreted
within that scope.

\section{Gaussianity of the PCA coefficients}
\label{sec:appendix_pca_gauss}
Our sampling method draws PCA coefficients from a multivariate Gaussian
fit to the training-set distribution (\S\ref{sec:methods}). Here we
quantify the residual departure from Gaussianity in each per-mode
coefficient distribution and check that it is small enough for that
generative model to be a useful approximation across the
(\eb,\qb) training grid.

For each binary in the training set, we project the concatenated
accretion-rate windows onto the global PCA basis and obtain coefficient
samples $c_k$ for modes $k \in [1, K]$. For each mode separately, we
standardize the coefficients,
\begin{equation}
    z_k = \frac{c_k - \langle c_k \rangle}{\sigma(c_k)},
\end{equation}
and quantify how far the resulting empirical distribution sits from
the standard normal $\mathcal{N}(0,1)$ using the first Wasserstein
distance \citep{Villani2009},
\begin{equation}
    W_{1,k} = \int_{-\infty}^{\infty} \left| F_{n,k}(z) - \Phi(z) \right| dz,
\end{equation}
where $F_{n,k}$ is the empirical CDF of the $n$ standardized samples
$\{z_k^{(i)}\}_{i=1}^n$ for that binary ($n = 500$ training windows
per mode), and $\Phi$ is the CDF of $\mathcal{N}(0,1)$. We evaluate
this as the mean absolute QQ-plot residual,
\begin{equation}
    W_{1,k} \approx \frac{1}{n} \sum_{i=1}^{n}
    \left| z_k^{(i)} - \Phi^{-1}\!\left(\tfrac{i - 1/2}{n}\right) \right|,
\end{equation}
in units of $z$ (standardized standard deviations). Unlike formal
goodness-of-fit test statistics (e.g.\ Anderson--Darling or
Kolmogorov--Smirnov), which scale with $n$ under any non-Gaussian
alternative so as to drive the test toward rejection at large
sample sizes, $W_{1,k}$ does not amplify with $n$: under any fixed
underlying distribution it converges to the true distance between
that distribution and $\mathcal{N}(0,1)$, so the absolute value of
$W_{1,k}$ remains directly interpretable as a $z$-unit displacement.

To obtain a single per-binary score, we aggregate $W_{1,k}$ over the
retained modes, weighted by the global PCA explained-variance ratios,
\begin{equation}
    \bar{W}_1 = \sum_{k=1}^{K} \tilde{\lambda}_k\, W_{1,k},
\end{equation}
where $\tilde{\lambda}_k$ are the explained-variance ratios
normalized over the retained modes, so that the modes contributing
most strongly to the global representation dominate the score.

\begin{figure}
	\centering
	\includegraphics[width=\columnwidth]{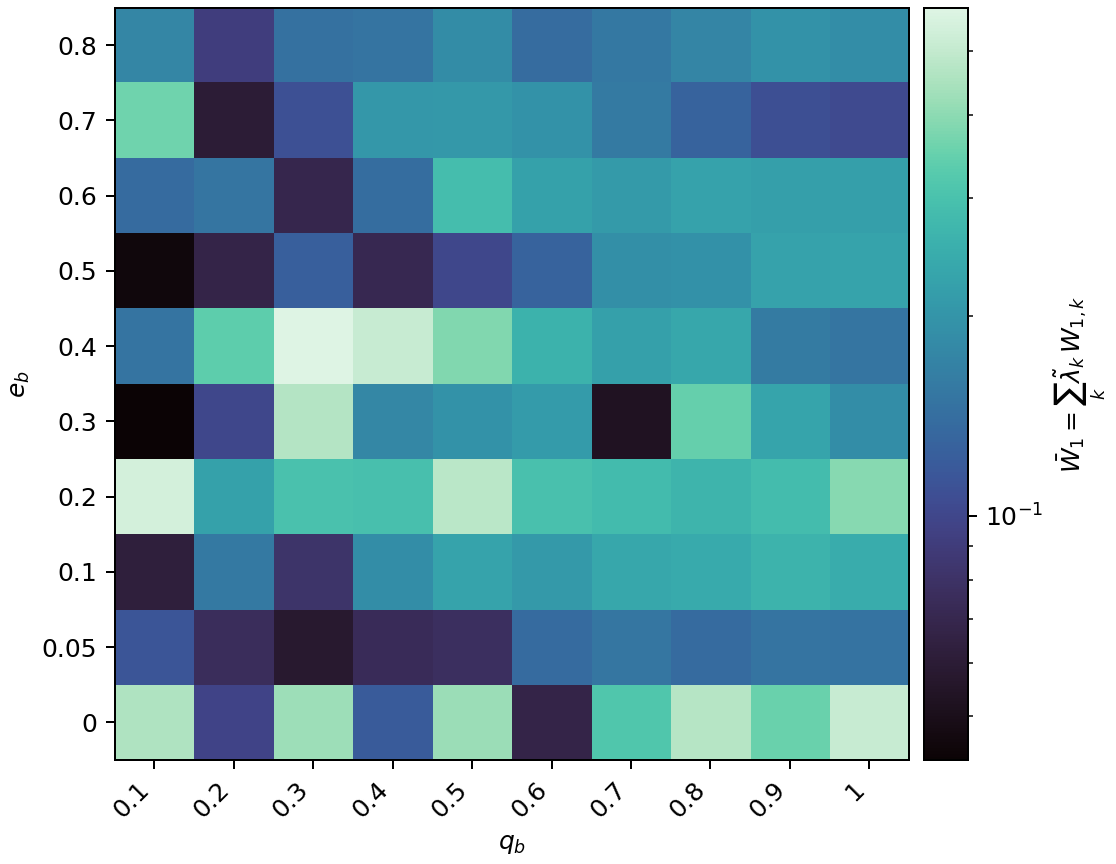}
	\caption{Per-binary variance-weighted Wasserstein-1 distance
	$\bar{W}_1 = \sum_k \tilde{\lambda}_k\, W_{1,k}$ between the
	standardized PCA coefficient distributions and $\mathcal{N}(0,1)$,
	across the (\eb,\qb) training grid; the horizontal axis is \qb\
	and the vertical axis is \eb. Values are in units of standardized
	standard deviations, so $\bar{W}_1 = X$ means the empirical CDFs
	differ from $\Phi$ by an average of $X$ $z$-units. Lower values
	(darker) indicate distributions closer to Gaussian.}
	\label{fig:appendix_pca_gauss}
\end{figure}

\begin{figure*}
	\centering
	\includegraphics[width=\textwidth]{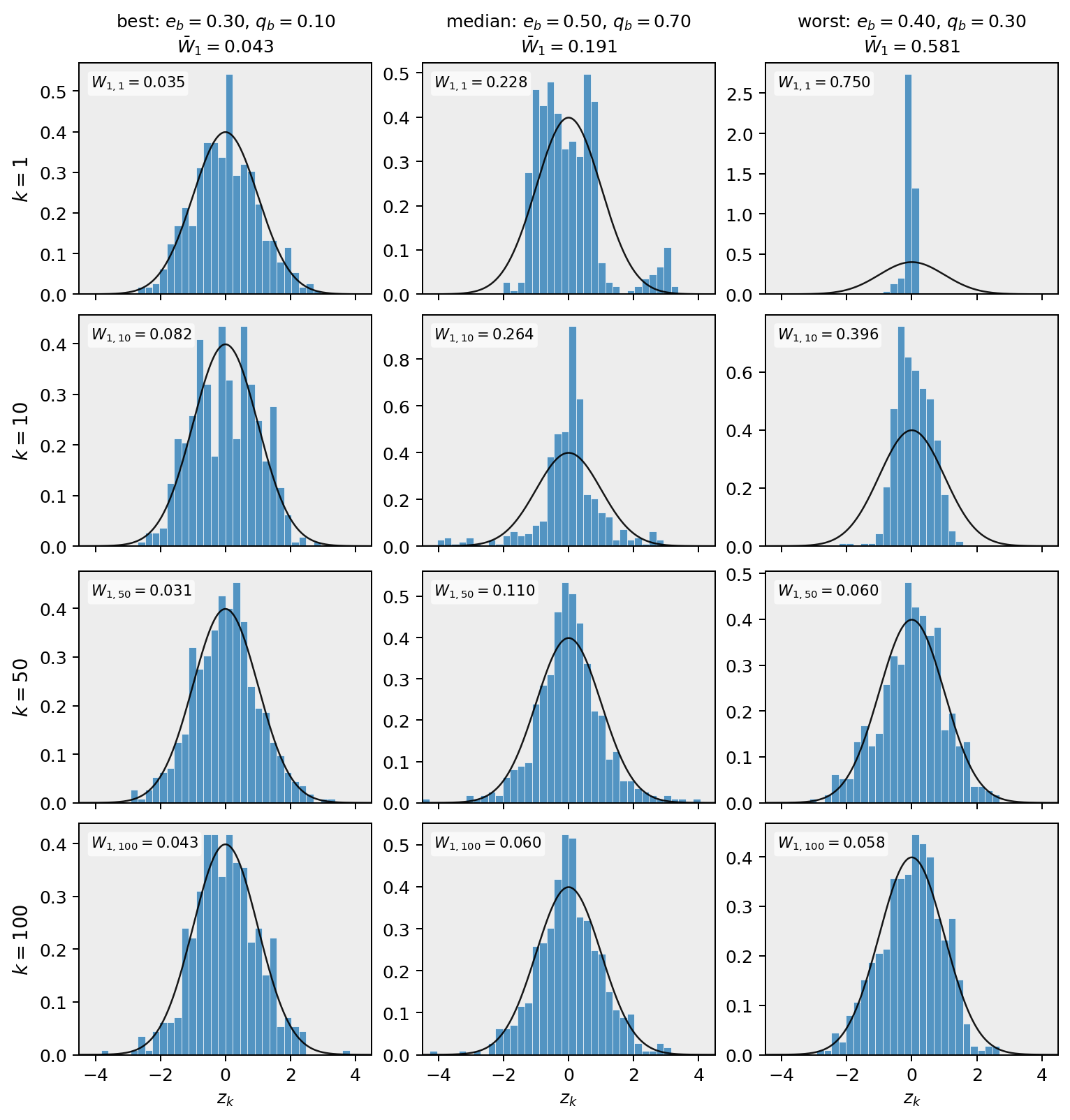}
	\caption{Standardized PCA coefficient distributions for three
	representative training cases (columns) and selected modes $k$
	(rows): from left to right, the binaries with the lowest, median,
	and highest $\bar{W}_1$ in Figure~\ref{fig:appendix_pca_gauss}.
	Each panel shows the empirical histogram of
	$z_k = (c_k - \langle c_k\rangle) / \sigma(c_k)$ overlaid with a
	standard normal $\mathcal{N}(0,1)$ reference; the per-mode
	Wasserstein-1 distance $W_{1,k}$ (in $z$-units) is printed in the
	upper-left of each panel, and each column header gives the binary
	(\eb,\qb) and its $\bar{W}_1$. In the best case all four sampled
	modes track $\mathcal{N}(0,1)$ closely; in the worst case the
	dominant mode is over-concentrated near zero --- two narrow spikes
	with thin tails rather than a smooth bell --- but $z_{k \ge 50}$
	recovers a Gaussian shape. Even in the worst case the per-mode
	deviations are bounded ($W_{1,k} \lesssim 1$), supporting the use
	of the Gaussian sampler as a tractable approximation rather than
	an exact match.}
	\label{fig:appendix_pca_hist}
\end{figure*}

Figure \ref{fig:appendix_pca_gauss} shows the resulting heatmap
across the (\eb,\qb) training grid. The aggregate $\bar{W}_1$
ranges from a few percent to roughly $0.6$ standard deviations
across the training set, with the bulk of binaries in the
$0.05$--$0.2$ range and a small number of cases reaching the upper
end of that range. Figure~\ref{fig:appendix_pca_hist} shows the
underlying per-mode histograms for three representative cases
(best, median, and worst $\bar{W}_1$) at modes
$k \in \{1, 10, 50, 100\}$, with $W_{1,k}$ printed in each panel.
In the best and median cases, the empirical distributions overlay
$\mathcal{N}(0,1)$ closely across all four sampled modes; in the
worst case, $z_1$ is over-concentrated near zero (two narrow spikes
flanking the origin, with thin tails), while $z_{k \ge 10}$
recovers a Gaussian shape. The per-binary deviations from
$\mathcal{N}(0,1)$ stay well within a standard deviation across the
entire grid, which we take as ``good enough'' for our Gaussian
sampling assumption. We note that the worst-case binary identified
here, (\eb=0.4, \qb=0.3), sits adjacent to one of the largest
residual biases in the end-to-end emulator predictions
(Figure~\ref{fig:test_timeseries}, at \eb=0.4, \qb=0.25),
suggesting that the residual non-Gaussianity in this region of
parameter space may contribute to the bias observed there.

\twocolumn

\bibliographystyle{mnras}
\bibliography{mybib} 

\end{document}